\documentclass[prd,showpacs,eqsecnum,amsmath,amssymb]{revtex4}

\newcommand{\be}{\begin{equation}}
\newcommand{\ee}{\end{equation}}

\newcommand{\pa}{\partial}

\def\nl{\\ & \quad}
\def\nlq{\\ & \qquad}

\def\picSin{((\mathbf{P}_1 \times \mathbf{S}_1) \cdot \mathbf{n}_{1 2})}
\def\piicSin{((\mathbf{P}_2 \times \mathbf{S}_1) \cdot \mathbf{n}_{1 2})}
\def\piicSiin{((\mathbf{P}_2 \times \mathbf{S}_2) \cdot \mathbf{n}_{1 2})}
\def\picSiin{((\mathbf{P}_1 \times \mathbf{S}_2) \cdot \mathbf{n}_{1 2})}
\def\picSipii{((\mathbf{P}_1 \times \mathbf{S}_1) \cdot \mathbf{P}_2)}
\def\SiSii{(\mathbf{S}_1 \cdot \mathbf{S}_2)}
\def\pipii{(\mathbf{P}_1 \cdot \mathbf{P}_2)}
\def\Sipi{(\mathbf{S}_1 \cdot \mathbf{P}_1)}
\def\Siipii{(\mathbf{S}_2 \cdot \mathbf{P}_2)}
\def\Sipii{(\mathbf{S}_1 \cdot \mathbf{P}_2)}
\def\Siipi{(\mathbf{S}_2 \cdot \mathbf{P}_1)}
\def\Sin{(\mathbf{S}_1 \cdot \mathbf{n}_{1 2})}
\def\Siin{(\mathbf{S}_2 \cdot \mathbf{n}_{1 2})}
\def\pin{(\mathbf{P}_1 \cdot \mathbf{n}_{1 2})}
\def\piin{(\mathbf{P}_2 \cdot \mathbf{n}_{1 2})}
\def\pipi{\mathbf{P}_1^2}
\def\piipii{\mathbf{P}_2^2}

\DeclareMathOperator{\Order}{\mathcal{O}}

\allowdisplaybreaks

\begin{document}

\title{ADM canonical formalism for gravitating spinning objects}

\author{Jan Steinhoff, Gerhard Sch\"afer, and Steven Hergt}
\affiliation{Theoretisch-Physikalisches Institut,
Friedrich-Schiller-Universit\"at,
Max-Wien-Pl.\ 1, 07743 Jena, Germany}

\date{\today}

\begin{abstract}
In general relativity, systems of spinning classical particles are implemented
into the canonical formalism of Arnowitt, Deser, and Misner
\cite{ADM62}. The implementation is made with the aid of
a symmetric stress-energy tensor and not a 4-dimensional
covariant action functional. The formalism is valid to terms linear in the
single spin variables and up to and including
the next-to-leading order approximation in the gravitational
spin-interaction part. The field-source
terms for the spinning particles occurring in the Hamiltonian are obtained from their expressions in
Minkowski space with canonical variables through 3-dimensional covariant
generalizations as well as from a suitable shift of projections of the curved spacetime stress-energy
tensor originally given within covariant spin supplementary
conditions. The applied coordinate conditions are the generalized
isotropic ones introduced by Arnowitt, Deser, and Misner.
As applications, the Hamiltonian of two spinning compact bodies with next-to-leading
order gravitational spin-orbit coupling, recently obtained by Damour, Jaranowski,
and Sch\"afer \cite{DJS07}, is rederived and the derivation of the
next-to-leading order gravitational spin(1)-spin(2) Hamiltonian, shown for the first time
in \cite{SHS07}, is presented.
\end{abstract}

\vspace{2ex}
\noindent

\maketitle

\vspace{2ex}

\section{Introduction}

Full implementation into canonical formalisms of general relativity (GR) and applications
have so far found only classical point masses \cite{ADM62, K61, KT72, OOKH73, OOK75, OK88, S85,
DS85, DS88, JS97, JS98, JS99, DJS01},
fluids \cite{KN84, H85, BMW85, BDS90, S90},
massive scalar fields \cite{S63a, S63c}, and gauge spin-1 fields,
including Maxwell \cite{ADM62, S63c} and Yang-Mills \cite{T80}.
The canonical implementation of spin-$\frac{1}{2}$ Dirac fields has been
performed in \cite{D62, K63, DI76, GH77, NT78}. Formally showing derivative coupling to the
spacetime metric, the Dirac fields resemble to the
classical spinning objects (pole-dipole particles) treated in our paper.
Problems of canonical gravity with derivative-coupled sources
are discussed in the comprehensive review by Isenberg and Nester \cite{IN80}.
Another common feature of Dirac fields and classical spinning objects is
the occurrence of surface terms in the Minkowski space algebra of the
stress-energy tensor, see Appendix \ref{flat_algebra}. The
canonical formulation of Dirac fields coupled to gravity is therefore a
valuable guide for the considerations in this paper and for future work.

Regarding classical pole-dipole particles in GR, see, e.g.,
\cite{P51, T59, D79}, the theory of special relativity (SR) tells us
that only for specific spin supplementary conditions (SSC), namely, the
Newton-Wigner (NW) ones \cite{NW49}, canonical
variables can be achieved. Related with a SSC is an implicit association of
the used coordinates to a specific center for
the particle: in case of the canonical NW SSC the center is called
center-of-spin, in case of the noncanonical
noncovariant M{\o}ller (in SR) or Corinaldesi-Papapetrou (in GR) SSC
\cite{M49, CP51} center-of-mass or
center-of-energy, and in case of the covariant
Fokker-Synge-Pryce (in SR) or Tulczyjew (in GR) SSC, \cite{F29, S35,
P48, T59}, center-of-inertia, see,
e.g., \cite{F65}. If one is interested in a theory with terms linear in
spin only, the Fokker-Synge-Pryce-Tulczyjew
SSC are identical with the Lanczos (in SR) \cite{L29} or Mathisson and
Pirani (in GR) SSC, \cite{M37, P56},
for history see, e.g., \cite{S99}.

In this paper, the canonical formalism by Arnowitt, Deser, and Misner (ADM), see \cite{ADM62},
will be applied to put the GR-dynamics of pole-dipole particles
into canonical form. The starting point will
not be a covariant action functional but rather the symmetric stress-energy tensor
of pole-dipole particles. The developed formalism is valid to terms
linear in spin and, in post-Newtonian framework, to
next-to-leading order approximation in the spin interaction part.
The formalism is applied to the derivations of the ADM Hamiltonian of two
spinning compact bodies with next-to-leading order gravitational spin-orbit coupling 
recently obtained in \cite{DJS07} and to the calculation of
the next-to-leading order gravitational spin(1)-spin(2)
Hamiltonian. The outcome of the latter calculation has been announced in
\cite{SHS07} already. It is hoped to develop the canonical formalism
to higher orders in future.

The canonical dynamics is only given in a reduced form in this paper,
i.e. all gauge degrees of freedom due to general coordinate invariance
are already fixed. A similar reduced formulation for gravitating Dirac fields
is given in \cite{K63}. The gauge independent canonical formalism for Dirac fields
is achieved for tetrad gravity instead of metric gravity in \cite{NT78}, i.e.
the vierbein instead of the metric is the fundamental dynamical variable.
An analogous canonical theory for classical spinning objects would be very
desirable. Of course, other methods are also well suited to incorporate spin effects into
the post-Newtonian expansion of general relativity \cite{KWW93,FBB06,GR06}. However,
a common problem of all formalisms, if one aims at a Hamiltonian formulation,
is to get Poisson brackets for the variables, or to find variables
that allow for standard Poisson brackets. Here the ADM formalism
presents itself as valuable because one is always close to the exact canonical
formulation of point-masses and the connection to the global
Poincar\'e algebra has already been studied in detail in the literature,
see, e.g., \cite{RT74,HRT76}. The global Poincar\'e algebra seems to be the
smartest tool to construct or validate Poisson brackets within
a post-Newtonian setting. Regarding interaction terms nonlinear in spin, the
ADM formalism has been proven useful too. Beyond leading order,
various new non-linear-in-spin binary Hamiltonians have been derived
recently, \cite{HS07}. For sake of completeness it should be mentioned that in \cite{YB93} a
covariant action functional approach to the dynamics of pole-dipole
particles in external gravitational fields has been introduced in
Routhian form using vierbein fields and in \cite{K72} the same
dynamics has been treated within the language of forms. A Lagrangian
approach is presented in \cite{N07}. In neither of the latter cases dynamical
canonical gravity has been envisaged.

It is important to point out that we will count post-Newtonian
orders, i.e. orders in $c^{-2}$, only in terms of velocity of light $c$
originally present in the Einstein field equations. Then both linear
momentum and spin are counted of the order zero. The next-to-leading
order in the spin interaction part therefore appears at the
second post-Newtonian order in this paper. This makes no statement about the numerical
value of these contributions, which can, of course, be much smaller compared to the second
post-Newtonian point-mass contributions (depending on the numerical values of the
spin variables). Some papers already respect
in their post-Newtonian expansion that the numerical value of the spin variables is assumed to
be of the order  $c^{-1}$. Then the next-to-leading order spin-orbit and spin-spin contributions, both second
post-Newtonian in our way of counting, are referred to as second-and-a-half and third
post-Newtonian contributions, respectively.

The paper is organized as follows. In Sec. II, the structure of the
ADM formalism is outlined. Emphasis
is put on the role the stress-energy tensor of the matter source of the Einstein
field equations plays in the ADM formalism.  In Sec. III, the matter
Hamiltonian and its relation  to the covariant 3-space components of the
matter stress-energy tensor are discussed.  The Sec. IV is devoted to
the stress-energy tensor of pole-dipole particles in Minkowski space in
canonical variables. The components of the stress-energy tensor
occurring in the curved spacetime Hamiltonian are constructed by 3-dimensional 
covariant generalization. The canonical linear momentum is identified as
generator of the global Poincar\'e algebra.  The action functional for
center-of-mass and spin motions is given. The Sec. V shows how the
components of the stress-energy tensor in the Hamiltonian can be
directly obtained in curved spacetime. In Sec. VI, consistency of the
obtained formalism is proved to the approximation of the Einstein field
equations treated in the paper. The Sec. VII is devoted to
applications. The next-to-leading order gravitational spin-orbit and
spin(1)-spin(2) Hamiltonians are calculated. In Sec. VIII, an
independent derivation of the next-to-leading order gravitational
spin(1)-spin(2) Hamiltonian is given using the lapse and shift functions
which are not involved in the calculation of the ADM Hamiltonians
of Sec. VII. Finally in Sec. IX, the Poincar\'e algebra is shown to
hold to the order of approximation of the developed formalism.
The Appendix A presents the local stress-energy tensor algebra for
pole-dipole particles in Minkowski space and the Appendix B gives the local
stress-energy tensor algebra in curved spacetime for nonspinning particles.
The Appendix C shows the applied regularization techniques.

Our units are $c=1$ and $G=1$, where $G$ is the Newtonian
gravitational constant. Greek indices will run over $0,1,2,3$, Latin over
$1,2,3$. For the signature of spacetime we choose +2. The short-cut notation
$ab$ ($= a^{\mu}b_{\mu} = a_{\mu}b^{\mu}$) for the scalar product of two vectors
$a^{\mu}$ and $b^{\mu}$ will be used. Round brackets denote index symmetrization,
i.e., $a^{( \mu} b^{\nu )} = \frac{1}{2} (a^{\mu} b^{\nu} + a^{\nu} b^{\mu})$.
The spatial part of a 4-vector $x$ is $\mathbf{x}$.

\section{Structure of the ADM formalism}

Crucial to the ADM formalism \cite{ADM62} is the Hamiltonian which generates the
full Einstein field equations, both the four
constraint equations and the 12 first order evolution
equations for the 3-metric $\gamma_{ij}$ and its
canonical conjugate $\frac{1}{16\pi} \pi^{ij}$, also see \cite{DW67, RT74},
\be
H = \int d^3x (N{\cal{H}} - N^i{\cal{H}}_i) + E[\gamma_{ij}] \,,
\ee
where $N$ and $N^i$ denote lapse and shift functions, which are merely
Lagrange multipliers. The super-Hamiltonian ${\cal{H}}$ and the supermomentum
${\cal{H}}_i$ densities decompose into gravitational field and matter parts as follows,
\be
{\cal{H}}={\cal{H}}^{\rm field} + {\cal{H}}^{\rm matter} \,, \qquad
{\cal{H}}_i={\cal{H}}^{\rm field}_i + {\cal{H}}^{\rm matter}_i \,,
\ee
where the field parts are given by
\be\label{Hfield}
{\cal{H}}^{\rm field} = - \frac{1}{16\pi\sqrt{\gamma}} \left[ \gamma \text{R}
	+ \frac{1}{2} \left( \gamma_{ij} \pi^{ij} \right)^2
	- \gamma_{ij} \gamma_{k l} \pi^{ik} \pi^{jl}\right] \, , \qquad
{\cal{H}}^{\rm field}_i = \frac{1}{8\pi} \gamma_{ij} \pi^{jk}_{~~ ; k} \, .
\ee
Here $\gamma$ is the
determinant of the 3-metric $\gamma_{ij} = g_{ij}$ of the spacelike
hypersurfaces $t = \text{const.}$, whereas the determinant of the
4-dim.\ metric $g_{\mu\nu}$ will be denoted $g$.
The canonical conjugate to $\gamma_{ij}$ is $\frac{1}{16\pi} \pi^{ij}$.
$\text{R}$ is the Ricci-scalar of the spacelike hypersurfaces and $;$
denotes the 3-dim.\ covariant derivative. The expressions for lapse and shift
functions are then $N=(-g^{00})^{-1/2}$ and $N^i=\gamma^{ij}g_{0j}$.
For simplicity, we will assume that the relation between field momentum $\pi^{ij}$ and
extrinsic curvature $K_{ij}$ is the same as in the vacuum case:
\begin{equation}\label{def_pi}
	\pi^{ij} = - \sqrt{\gamma} (\gamma^{ik}\gamma^{jl} - \gamma^{ij}\gamma^{kl})K_{kl}
\end{equation}
The energy $E$ is defined by
\be\label{ADM_energy}
E = \frac{1}{16\pi}\oint d^2 s_i (\gamma_{ij,j} -\gamma_{jj,i}) \, ,
\ee
where , denotes partial space derivatives and  $d^2 s_i$ the
2-dim.\ spatial volume element at spatial infinity.
The surface expression $E$ makes the Hamilton variational principle well-defined
also for variations which do not have compact support. After imposing coordinate
conditions and constraint equations,
\be\label{const_eq}
\frac{\delta H}{\delta N} \equiv {\cal{H}}=0, \qquad - \frac{\delta
H}{\delta N^i} \equiv {\cal{H}}_i = 0 \, ,
\ee
$E$ turns into the ADM Hamiltonian $H_{\rm ADM}$.
Comparing these constraints with the Einstein equations, projected onto
the spacelike hypersurfaces, results in
\be\label{Hmatter}
{\cal{H}}^{\rm matter}= \sqrt{\gamma}T^{\mu\nu}n_{\mu}n_{\nu} =
N\sqrt{-g}T^{00} \, , \qquad
{\cal{H}}^{\rm matter}_i= - \sqrt{\gamma}T^{\nu}_i n_{\nu} =
\sqrt{-g}T^{0}_i \, ,
\ee
where $\sqrt{-g}T^{\mu\nu}$ is the stress-energy tensor density of the
matter system. The timelike unit 4-vector $n_{\mu} = (-N,0,0,0)$ points
orthogonal to the spacelike hypersurfaces.
The evolution equations of the field, before imposing constraints and
coordinate conditions, read
\be\label{evol_eq}
\frac{1}{16\pi} \frac{\pa {\pi}^{ij}}{\pa t} =
	- \frac{\delta H}{\delta \gamma_{ij}} \, , \qquad
\frac{1}{16\pi} \frac{\pa \gamma_{ij}}{\pa t} =
	\frac{\delta H}{\delta \pi^{ij}} \, .
\ee
The coordinate conditions must be preserved under this time evolution.
These additional constraints fixate lapse and shift functions.

The ADMTT gauge \cite{ADM62}, being the most often used and best adapted coordinate
condition for explicit calculations, is given by
\be\label{cc}
\gamma_{ij} = \psi^4 \delta_{ij} + h^{\text{TT}}_{ij}\,, \quad \text{or} \quad
3\gamma_{ij,j} - \gamma_{jj,i} = 0 \,, \quad \text{and} \quad
\pi^{ii} = 0 \,.
\ee
Here $h^{\text{TT}}_{ij}$ has the properties  $h^{\text{TT}}_{ii} = h^{\text{TT}}_{ij,j}=0$
(transverse, traceless). After imposing the constraint equations (\ref{const_eq}),
the remaining 4 (reduced) field equations read
\be\label{red_evol_eq}
\frac{1}{16\pi} \frac{\pa {\pi}_{\text{TT}}^{ij}}{\pa t} =
	- \frac{\delta H_{\rm ADM}}{\delta h^{\text{TT}}_{ij}} \, ,  \qquad
\frac{1}{16\pi} \frac{\pa h^{\text{TT}}_{ij}}{\pa t} =
	\frac{\delta H_{\rm ADM}}{\delta \pi_{\text{TT}}^{ij}} \, ,
\ee
where $\pi_{\text{TT}}^{ij}$ denotes the transverse traceless part of $\pi^{ij}$
and the variational derivatives must include a projection onto the
transverse traceless part. We will call the phase space consisting of
$h^{\text{TT}}_{ij}$, $\frac{1}{16\pi}{\pi}_{\text{TT}}^{ij}$ and canonical matter variables
the reduced phase space, whereas the nonreduced phase space consists of
$\gamma_{ij}$, $\frac{1}{16\pi} \pi^{ij}$ and canonical matter variables.

The fundamental problem to be solved are the forms of the super-Hamiltonian
and supermomentum densities for pole-dipole particles in canonical
variables. Our approach will be as follows.
We first construct the stress-energy tensor in Minkowski space with canonical variables. Then
taking into account that, respectively, ${\cal{H}}^{\rm matter}$ and ${\cal{H}}^{\rm matter}_i$ are
scalar and covariant vector densities with respect to 3-dim.\ coordinate transformations, we put
these expressions into 3-dim.\ covariant forms (this procedure had been
suggested already by Boulware and Deser \cite{BD67}). Afterwards we show
that the same result can be obtained by Lie-shifting certain components
of the stress-energy tensor with rotational-free parallel transport of
the linear momentum fields.

\section{Consistency conditions}

The Hamilton variational principle must generate the Einstein equations. This
trivial fact leads to several consistency conditions for the matter part
of the Hamiltonian,
\be
H^{\rm matter} = \int d^3x (N{\cal{H}}^{\rm matter} - N^i{\cal{H}}_i^{\rm matter})\,.
\ee
Lapse and shift are Lagrange multipliers, so ${\cal{H}}^{\rm matter}$ and
${\cal{H}}^{\rm matter}_i$ must be independent of them.
Equation (\ref{Hmatter}) then already ensures that the constraint Eqs.\ (\ref{const_eq})
are correct. The evolution Eqs.\ (\ref{evol_eq}) coincide with the
Einstein equations if and only if
\begin{align}
\frac{\delta H^{\rm matter}}{\delta \pi^{ij}} &= 0 \label{global_cons2} \,, \\
\frac{\delta H^{\rm matter}}{\delta \gamma^{ij}} &=
\frac{1}{2} N \sqrt{\gamma} T_{ij} \,. \label{global_cons1}
\end{align}
Violation of the first condition would give an incorrect evolution equation for
$\gamma_{ij}$. This is critical, because the geometric meaning of this equation is the
\emph{definition} of the extrinsic curvature $K_{ij} \equiv - N \Gamma^0_{ij}$, additional
terms here imply leaving Riemannian geometry. This might be fixed by adjusting
Eq.\ (\ref{def_pi}), see \cite{K77}. The second condition ensures that the evolution equation
for $\pi^{ij}$ fits with the Einstein field equations.

The first condition, Eq.\ (\ref{global_cons2}), is equivalent to the local equations
\be\label{local_cons2}
\frac{\delta {\cal{H}}^{\rm matter}(\mathbf{x})}{\delta \pi^{ij}(\mathbf{x}')} = 0 \,, \qquad
\frac{\delta {\cal{H}}_k^{\rm matter}(\mathbf{x})}{\delta \pi^{ij}(\mathbf{x}')} = 0 \,.
\ee
The second condition, Eq.\ (\ref{global_cons1}),  then implies that also $T_{ij}$ is independent of $\pi^{ij}$.
If and only if $T_{ij}$ does also not depend on lapse and shift, then the local version of
the second condition reads
\be\label{local_cons1}
\frac{\delta {\cal{H}}^{\rm matter}(\mathbf{x})}{\delta \gamma^{ij}(\mathbf{x}')} =
	\frac{1}{2} \sqrt{\gamma} T_{ij}(\mathbf{x}) \delta(\mathbf{x} - \mathbf{x}') \,, \qquad
\frac{\delta {\cal{H}}_k^{\rm matter}(\mathbf{x})}{\delta \gamma^{ij}(\mathbf{x}')} = 0 \,.
\ee
In Appendix \ref{curved_algebra} it will be shown that Eq.\ (\ref{local_cons1}) is equivalent to the simple
constraint algebra (\ref{diffeo1}\nobreakdash--\ref{diffeo3}).
The conditions given in Eq.\ (\ref{local_cons1}) are very restrictive, as they imply
that ${\cal{H}}^{\rm matter}$ cannot depend on derivatives of $\gamma^{ij}$, and
${\cal{H}}_k^{\rm matter}$ cannot depend on $\gamma^{ij}$ at all.
Together with (\ref{def_pi}), this defines a kind of simple coupling of matter
to gravity. Gravitating classical spinning objects and Dirac fields are not of
this kind. However, our canonical formulation of spinning objects will exactly fulfill
(\ref{def_pi}) and (\ref{local_cons2}), and also at least approximately
(\ref{global_cons1}), see Sec.\ \ref{consistency}.

None of the preceding consistency conditions validates the canonical matter variables directly,
in our case position, linear momentum and spin. In a theory that is of the simple kind mentioned above,
this can be done via a local algebra for $\mathcal{H}^{\rm matter}$ and $\mathcal{H}^{\rm matter}_i$
on the nonreduced phase space, Eqs.\ (\ref{matter_algebra_1}\nobreakdash--\ref{matter_algebra_3}).
We will instead consider the global Poincar\'e algebra, which is a consequence of the
asymptotic flatness and is represented by Poisson-brackets of the corresponding conserved
quantities. So besides the ADM energy (\ref{ADM_energy}) also total linear momentum $P_i$,
total angular momentum $J_i\equiv \frac{1}{2} \epsilon_{ijk} J_{jk}$ and the boost vector
$K^i$ are conserved and given by surface integrals at spatial infinity. The boosts have an
explicit dependence on the time $t$ and can be decomposed as $K^i \equiv G^i - t P_i$, where
$X^i \equiv G^i / E$ is the coordinate of the center-of-mass. $G^i$ will be called center-of-mass
vector in the following. The corresponding surface integrals read, with spatial coordinates denoted $x^i$:
\begin{gather}
	P_i = - \frac{1}{8\pi}\oint d^2 s_k \pi^{ik} \,, \qquad
	J_{ij} = - \frac{1}{8\pi}\oint d^2 s_k ( x^i \pi^{jk} - x^j \pi^{ik}) \,, \\
	G^i = \frac{1}{16\pi}\oint d^2 s_k \left[ x^i ( \gamma_{kl,l} - \gamma_{ll,k} )
		- \gamma_{ik} + \delta^i_k \gamma_{ll} \right]\,.
\end{gather}
After imposing constraints and coordinate conditions, these quantities have well-defined Poisson-brackets on 
the reduced phase-space \cite{RT74}, and the Poincar\'e algebra can be verified.
At the second post-Newtonian level for spin, and also in the spatial conformally flat case
$\gamma_{ij}=\psi^4\delta_{ij}$, we have, by virtue
of the momentum constraints ${\cal{H}}_i=0$ and the ADMTT gauge, the following simple expressions:
\be\label{conserved}
	P_i = \int d^3 x \, \mathcal{H}_i^{\rm matter} \,, \qquad
	J_{ij} = \int d^3 x \, ( x^i \mathcal{H}_j^{\rm matter} - x^j \mathcal{H}_i^{\rm matter})\,.
\ee
In the ADMTT gauge it also holds:
\be\label{conserved2}
	E = H_{\rm ADM} = - \frac{1}{2 \pi} \int{ d^3 x \, \Delta \psi } \,, \qquad
	G^i = - \frac{1}{2 \pi} \int{ d^3 x \, x^i \Delta \psi }\,.
\ee
After solving the Hamilton constraint $\mathcal{H}=0$, $\psi$ can be expressed in terms of
canonical variables of the reduced phase space, and the Poincar\'e algebra can be verified.

A final remark concerns the canonical spin variables. Imposing the standard Poisson-bracket
algebra of angular momentum for the spin variables, it is clear that the squared euclidean
length of the spin, being a Casimir operator, will commute with all other canonical variables.
Therefore this length is constant in time, as it will also commute with the Hamiltonian.

\section{Pole-dipole particle stress-energy tensor in canonical variables}

Calculating $\mathcal{H}^{\rm matter}$ and $\mathcal{H}^{\rm matter}_i$ via
(\ref{Hmatter}) in the Minkowskian case and then going over to their
3-dim.\ covariant generalizations has the advantage that $\mathcal{H}^{\rm matter}$
and $\mathcal{H}^{\rm matter}_i$ will definitely not depend on lapse, shift, and
$\pi^{ij}$ or $K^{ij}$. This is a serious problem when working in curved
spacetime. Then our matter variables (in particular, spin and momentum of the particles,
but not their position) have to be redefined to suit the consistency conditions of the previous
section. It was also observed in \cite{BD67} that the correct general relativistic source terms
in the constraint equations for low spin ($\le 1$) fields, including
electrodynamics, can be achieved by expressing their flat-space action
in a 3-dim.\ covariant form, and redefining canonical variables
in a way that leaves them unchanged in the flat case. This is similar to
our approach. In the next Section we will show that a curved spacetime approach
is also possible, yielding the same result as in the present Section.

Because of its importance for later transition to curved spacetime with
canonical variables, the stress-energy tensor density for an electric charge-free
pole-dipole particle in curved spacetime takes the form, to linear order in spin,
see, e.g., \cite{T02},
\begin{align}\label{stress_energy}
\sqrt{-g} \ T^{\mu\nu} &= \int d\tau \left[mu^{\mu}u^{\nu} \delta_{(4)}
-  (S^{\alpha (\mu}  u^{\nu)}\delta_{(4)})_{||\alpha}\right] \\
&= p^{\mu}v^{\nu} \delta - (S^{\alpha (\mu}
v^{\nu)}\delta)_{,\alpha} - S^{\alpha
(\mu}\Gamma^{\nu)}_{\alpha\beta}v^{\beta}\delta \, ,
\end{align}
applying the Tulczyjew SSC or, equivalently, the Mathisson-Pirani SSC
\be
S^{\mu\nu}u_{\nu}=0\,.
\ee
Here $v^{\mu} = u^{\mu}/u^0$ and $p^{\mu} = m u^{\mu}$,
particularly $p_i = mu_i$, with mass $m$ and $g_{\mu\nu}u^{\mu}u^{\nu} = -1$.
The Christoffel symbols are denoted $\Gamma^{\lambda}_{\mu\nu}$ as usual,
and the 4-dim.\ covariant derivative by $||$. The
4-dim.\ spin tensor $S^{\mu\nu}$ has the property $S^{\mu\nu}= - S^{\nu\mu}$.
$\tau$ is a proper time parameter running from $-~\infty$ to + $\infty$ with $u^{\mu} =
\frac{dz^{\mu}}{d\tau}$, where $z^{\mu}$ is the 4-dim.\ position variable
of the particle. The coordinate time velocity of the particle, $v^{\mu}$, is identical
with $\frac{dz^{\mu}}{dt}$. The Dirac delta functions, $\delta_{(4)} \equiv \delta(x - z)$
and $\delta \equiv \delta(\mathbf{x} - \mathbf{z})$, are normalized such that
$\int d^4x \, \delta_{(4)} = \int d^3x \, \delta = 1$ holds.

Furthermore, again to leading order in spin, it holds
\be
\frac{DS^{\mu\nu}}{d\tau} = 0 \, ,
\ee
where $D$ denotes the 4-dim.\ covariant differential. Obviously,
\be
S^{\mu\nu}S_{\mu\nu} \equiv 2s^2 = \text{const}.
\ee
is valid.

The transition to Minkowski space results in the stress-energy tensor
\begin{align}
T^{\mu\nu} &= \int d\tau \left[mu^{\mu}u^{\nu} \delta_{(4)}
-  (S^{\alpha (\mu}  u^{\nu)}\delta_{(4)})_{,\alpha}\right]  \nonumber \\
&= p^{\mu}v^{\nu} \delta - (S^{\alpha (\mu} v^{\nu)}\delta)_{,\alpha} \, .
\end{align}
Now we proceed to the Newton-Wigner SSC in making the following shift
of the particle coordinates
\be\label{flat_center}
\hat{z}^{\mu} = z^{\mu} -  \frac{S^{\mu\nu}n_{\nu}}{m-np} \, ,
\ee
where $np \equiv n_{\mu} p^{\mu} = - \sqrt{m^2 + \gamma^{ij} p_i p_j}$,
as well as introducing the spin tensor $\hat{S}^{\mu\nu}$ by the
relation, see, e.g., \cite{F65},
\be\label{flat_spin_def}
{S}^{\mu\nu} = {\hat S}^{\mu\nu} 
+ p^{\mu} n_{\lambda} {\hat S}^{\nu\lambda} /m
- p^{\nu} n_{\lambda}  {\hat S}^{\mu\lambda} /m \, ,
\ee
which results in
\be\label{flat_SSC}
(n_{\nu} + p_{\nu}/m) {\hat S}^{\mu\nu}=0\,.
\ee
This turns the stress-energy tensor into the form
(from now on $\delta \equiv \delta(\mathbf{x} - \mathbf{\hat{z}})$)
\be\label{all_components}
\hat{T}^{\mu\nu}(x,\hat{z}) \equiv  T^{\mu\nu}(x,z) = p^{\mu}v^{\nu}
\delta - ({\hat S}^{\alpha (\mu}
v^{\nu)}\delta)_{,\alpha} \, ,
\ee
because $\dot{\hat{z}}^{\mu} = \dot{z}^{\mu}$ (dot means time
derivative) to linear order in spin.
The new spin tensor  $\hat{S}^{\mu\nu}$ has the important property that
\be
S^{\mu\nu}S_{\mu\nu} = \hat{S}_{ij}\hat{S}_{ij} = \text{const}.
\ee
is valid.

The components of the stress-energy tensor, relevant for the ADM formalism, read
\begin{align}
\sqrt{\gamma}\hat{T}^{\mu\nu}n_{\mu}n_{\nu} &= - np\delta  -
\left(\delta_{ij}\delta_{kl}\frac{p_l}{m-np}{\hat
S}_{jk}\delta\right)_{,i} , \\
- \sqrt{\gamma}\hat{T}^{\nu}_i n_{\nu} &= p_i\delta
+ \frac{1}{2} \left(\left(\delta_{mk}{\hat S}_{ik} -
(\delta_{mk}\delta_{ip} + \delta_{mp}
\delta_{ik})\delta_{ql}
{\hat S}_{qp}\frac{p_l p_k}{np (m-np)}\right)\delta\right)_{,m} .
\end{align}
These components of the stress-energy tensor fulfill the Poisson-bracket
algebra a stress-tensor has to fulfill in Minkowski space, see
\cite{BD67,S98}. Details are given in Appendix \ref{flat_algebra}.

The 3-dim.\ covariant generalizations of these expressions read
(; denotes the 3-dim.\ covariant derivative)
\begin{align}
\mathcal{H}^{\rm matter} &\equiv \sqrt{\gamma}\hat{T}^{\mu\nu}n_{\mu}n_{\nu} =
	- np\delta - \left(\gamma^{ij}\gamma^{kl}\frac{p_l}{m-np}{\hat S}_{jk}\delta\right)_{,i}
	\equiv  N^2 \sqrt{\gamma} \hat{T}^{00} \, , \label{H} \\
\mathcal{H}_i^{\rm matter} &\equiv - \sqrt{\gamma}\hat{T}^{\nu}_i n_{\nu} =
	p_i\delta + \frac{1}{2} \left(\left(\gamma^{mk}{\hat S}_{ik}
	- (\gamma^{mk}\delta_i^p + \gamma^{mp} \delta_i^k)\gamma^{ql} {\hat S}_{qp}
		\frac{p_lp_k}{np (m-np)}\right)\delta\right)_{;m}
	\equiv N\sqrt{\gamma}\hat{T}^0_i \, . \label{Hi}
\end{align}
Correspondingly,
\be
\gamma^{ik}\gamma^{jl}\hat{S}_{ij}\hat{S}_{kl} = 2 s^2 = \text{const}.
\ee
has to hold. The new canonical spin variables $S_{(i)(j)}$ (the round
brackets make allusion to implicit dreibein components)
are defined such that
\be
\gamma^{ik}\gamma^{jl}\hat{S}_{ij}\hat{S}_{kl}=S_{(i)(j)}S_{(i)(j)}=2s^2
\ee
is valid. This can be achieved by constructing $e_{ij}$ as the symmetric
matrix square root of symmetric $\gamma_{ij}$ ($\gamma_{ij}=\gamma_{ji}$),
\be
e_{il}e_{lj}=\gamma_{ij} \, , \qquad e_{ij}=e_{ji} \, .
\ee
Then it holds
\be
\hat{S}_{kl}=e_{ki}e_{lj} S_{(i)(j)} \, .
\ee
The condition $e_{ij}=e_{ji}$ had also been imposed on the spatial part of the vierbein
field in \cite{K63} in order to achieve a canonical formalism for the spin-$\frac{1}{2}$ field.

If the 3-metric is represented in the form
\be
\gamma_{ij} = \delta_{ij}+ h_{ij}\,, \quad | h_{ij}| << 1\,,
\ee
the solution for $e_{ij}$ reads (with some abuse of notation)
\be
e_{ij} = \sqrt{\delta_{ij}+ h_{ij}} = \delta_{ij}
+ \frac{1}{2} h_{ij} - \frac{1}{8} h_{ik}h_{kj} + \ldots
\ee
and the variation of $e_{ij}$ is given by  
\be
\delta e_{ij} = \frac{1}{2} \delta \gamma_{ij} - \frac{1}{8} (h_{kj}
 \delta\gamma_{ik} + h_{ik} \delta \gamma_{kj})  + \ldots \; .
\ee
It may be pointed out that the simple variational relation for dreibein
fields $e_{(j)i}$, where $\gamma_{ij} = e_{(k)i} e_{(k)j}$, of the
form $2\delta e_{(j)i} = e_{(j)k} \gamma^{kl}\delta \gamma_{li}$
is not valid for our symmetric matrix square root in general; exceptions 
are isotropic metrics.

Recalling Eq.\ (\ref{conserved}), the new canonical momentum $P_i$ is defined
in the way that the following structure holds,
\be\label{Hicanon}
\mathcal{H}_i^{\rm matter} = P_i\delta
+ \frac{1}{2} \left[\left(\gamma^{mk}{\hat S}_{ik} -\frac{P_lP_k}{nP (m-nP)}
(\gamma^{mk}\delta_i^p + \gamma^{mp} \delta_i^k)\gamma^{ql}
{\hat S}_{qp}\right)\delta\right]_{,m} ,
\ee
where
\be\label{def_P}
P_i \equiv p_i - \frac{1}{2}
\left[\gamma^{lj}\gamma^{kp}\gamma_{il,p} - \frac{p_mp_q}{np(m-np)}
\gamma^{mj}\gamma^{kl}\gamma^{qp}\gamma_{lp,i}\right]{\hat S}_{jk} \, .
\ee
Hereof, we get
\be\label{Hcanon}
\mathcal{H}^{\rm matter} = - nP \delta - \frac{1}{2} t_{ij}^{k} \gamma^{ij}_{~~,k}
-\left(\frac{P_l}{m-nP}\gamma^{ij}\gamma^{kl}
{\hat S}_{jk}\delta\right)_{,i} \, ,
\ee
where the quantity $t_{ij}^{k}$ can be related to the flat $\sqrt{\gamma}\hat{T}_{ij}$ via Eq.\ (\ref{all_components}):
\be\label{Tij}
	\sqrt{\gamma}\hat{T}_{ij} = - \frac{P_i P_j}{nP} \delta + t_{ij,k}^k + \Order{(G)}\,, \qquad
	t_{ij}^{k} \equiv \gamma^{kl}\frac{\hat{S}_{l(i} P_{j)}}{nP} \delta
		+ \gamma^{kl}\gamma^{mn}\frac{\hat{S}_{m(i} P_{j)} P_n P_l}{(nP)^2(m-nP)}
			\delta\,.
\ee

The crucial question now is for the canonical variables. For both the
second post-Newtonian order approximation for spin
and the spatial conformally flat case in general 
we get for linear and angular momentum
\begin{align}
	P_i &\equiv \int d^3 x \, \mathcal{H}_i^{\rm matter} = P_i\,, \label{momentum} \\
	J_{ij} &\equiv \int d^3 x \, ( x^i \mathcal{H}_j^{\rm matter} - x^j \mathcal{H}_i^{\rm matter})
		=\hat{z}^iP_j- \hat{z}^jP_i + S_{(i)(j)} \,.\label{amomentum}
\end{align}
It is important that these expressions were achieved in the ADMTT gauge and with $e_{ij}=e_{ji}$.
Under these conditions both generators of the global Poincar\'e group
fit with the standard Poisson-brackets,
\be
\{{\hat z}^i(t), P_j(t)\} = \delta_{ij} \, , \qquad \{S_{(i)}(t), S_{(j)}(t)\} =
\epsilon_{ijk} S_{(k)}(t) \, , \qquad \text{zero otherwise} \, ,
\ee
where $S_{(i)(j)} = \epsilon_{ijk}S_{(k)}$ ($S_{(i)}S_{(i)} = s^2$) with the completely
antisymmetric Levi-Civita tensor $\epsilon_{ijk}$ ($\epsilon_{123}=1$).
In the following we will also use the notations ${\bf S}$ for $S_{(i)}$,
${\bf P}$ for $P_i$, and ${\bf Z}$ for $\hat{z}^i$. The commutation relations of the
field variables still read
\be
\{h^{\text{TT}}_{ij}({\bf x},t), \pi_{\text{TT}}^{kl}({\bf x}',t)\}
= 16\pi \delta^{\text{TT}kl}_{ij}\delta({\bf x} - {\bf x}') \, , \qquad \text{zero otherwise} \, ,
\ee
where
\begin{align}
\begin{split}
\delta^{\text{TT}kl}_{ij} &\equiv \tfrac{1}{2} [(\delta_{il}-\Delta^{-1}\pa_{i}\pa_{l})(\delta_{jk}-\Delta^{-1}\pa_{j}\pa_{k})+
(\delta_{ik}-\Delta^{-1}\pa_{i}\pa_{k})(\delta_{jl}-\Delta^{-1}\pa_{j}\pa_{l}) \\
&\quad -(\delta_{kl}-\Delta^{-1}\pa_{k}\pa_{l})(\delta_{ij}-\Delta^{-1}\pa_{i}\pa_{j})] 
\end{split}
\end{align}
with the inverse Laplacian $\Delta^{-1}$ and the partial space-coordinate derivatives $\pa_i$.
Herewith we have completed the calculation of the source terms applicable to the
ADM formalism. Crucial for our approach is the property of our spin
variables ${\bf S}$ to have conserved euclidean length. Further discussion of the
consistency of our formalism is given in Sec. \ref{consistency}.

The ADM Hamiltonian, written for a many-particle system (numbering $a=1,2,...$) depends on
the following variables,
\be
H_{\rm ADM} = H_{\rm ADM}[{\hat z}^i_a, P_{ai}, S_{a(i)}, h^{\text{TT}}_{ij}, \pi_{\text{TT}}^{kl}]
\ee
and the corresponding action $W$ reads (dot means time derivative)
\be\label{action}
W= \int dt \left(\sum_a P_{ai} \dot{\hat z}_{a}^i + \sum_a S^{(i)}_a\Omega^{(i)}_a
	+ \frac{1}{16 \pi} \int d^3x \, \pi^{ij}_{\text{TT}}\dot{h}^{\text{TT}}_{ij}
	- H_{ADM}\left[ {\hat z}^i_a, P_{ai}, S^{(j)}_a, h^{\text{TT}}_{ij}, \pi^{ij}_{\text{TT}}\right]\right) ,
\ee
where $\Omega^{(i)}_a = \frac{1}{2}\epsilon_{ijk} \Lambda_{a(l)(j)}\dot{\Lambda}_{a(l)(k)}$
with $\Lambda_{a(i)(k)}\Lambda_{a(j)(k)}=\Lambda_{a(k)(i)}\Lambda_{a(k)(j)}=\delta_{ij}$.
Hereof, by variation of $W$ with respect to $P_{ai}$, ${\hat z}^i_a$, $S^{(i)}_a=\frac{1}{2}\epsilon^{ijk} S_{a(j)(k)}$,
$\Lambda_{a(i)(j)}$ in the forms $\delta P_{ai}$, $\delta {\hat z}^i_a$, $\delta
S^{(i)}_a$, $\delta \Theta^{(i)}_a =  \frac{1}{2}\epsilon_{ijk} \Lambda_{a(l)(j)} \delta
{\Lambda}_{a(l)(k)}$, the equations of motion follow:
\begin{alignat}{2}
\dot{\hat z}^i_a(t) &= \frac{\delta \int dt' H_{ADM}}{\delta P_{ai}(t)}\,, & \qquad
\dot{P}_{ai}(t) &= - \frac{\delta \int dt' H_{ADM}}{\delta {\hat z}^i_a(t)}\,, \\
\Omega^{(i)}_a(t) &= \frac{\delta \int dt' H_{ADM}}{\delta S^{(i)}_a(t)}\,, & \qquad
\dot{S}^{(i)}_a(t) &= \epsilon_{ijk} \Omega^{(j)}_a(t)S^{(k)}_a(t)\,.
\end{alignat}
The field evolution is obviously given by Eq.\ (\ref{red_evol_eq}).

Finally, the transition to a Routhian reads
\be\label{routh}
R[{\hat z}^i_a, P_{ai}, S_{a(k)}, h^{\text{TT}}_{ij}, \dot{h}^{\text{TT}}_{ij}] =
H_{\rm ADM}[{\hat z}^i_a, P_{ai}, S_{a(k)}, h^{\text{TT}}_{ij}, \pi_{\text{TT}}^{kl}]
- \frac{1}{16\pi} \int d^3x~\pi_{\text{TT}}^{kl}\dot{h}^{\text{TT}}_{kl} \,,
\ee
with the field equations
\be
\frac{\delta \int R(t')  dt'}{\delta h^{\rm \text{TT}}_{ij}(x^k,t)} = 0 \,,
\ee
and the equations of motion 
\begin{alignat}{2}
\dot{\hat z}^i_a(t) &= \frac{\delta  \int dt' R}{\delta P_{ai}(t)}\,, & \qquad
\dot{P}_{ai}(t) &= - \frac{\delta  \int dt' R}{\delta {\hat z}^i_a(t)}\,, \\
\Omega^{(i)}_a(t) &= \frac{\delta \int dt' R}{\delta S^{(i)}_a(t)}\,, & \qquad
\dot{S}^{(i)}_a(t) &= \epsilon_{ijk} \frac{\delta \int dt' R}{\delta S^{(j)}_a(t)} S^{(k)}_a(t)\,.
\end{alignat}
The Routhian is very suitable for the derivation of an autonomous,
conservative Hamiltonian for the matter, where the solution
$h^{\text{TT}}_{ij}$ of the field equations is replaced by the matter
variables, see \cite{JS98}.

\section{Spacetime approach to the stress-energy tensor in canonical variables\label{curved}}

The 3-dim.\ derivation in the previous section of the needed stress-energy components
does not show up which 4-dim.\ object in curved spacetime is behind the performed
construction. This will be clarified in this Section. Starting from our
original curved spacetime stress-energy tensor density with covariant SSC, Eq.\ (\ref{stress_energy}),
we add up the following Lie-shift to it,
\be\label{ohashi}
(\sqrt{-g}T^{\mu\nu})_{\rm shifted} \equiv \sqrt{-g}T^{\mu\nu} +
{\cal{L}}_{m^{\sigma}}\sqrt{-g}T^{\mu\nu} = \int d\tau \left[\left(mu^{\mu}u^{\nu} -
\frac{Dm^{(\mu}}{d\tau} u^{\nu)}\right)\delta_{(4)}
-  ({\hat S}^{\alpha (\mu}u^{\nu)}\delta_{(4)})_{||\alpha}\right],
\ee
where $m^{\nu} =  -  S^{\nu\mu}n_{\mu}/(1-nu)$ and
\begin{gather}
{S}^{\mu\nu} = {\hat S}^{\mu\nu} + u^{\mu} n_{\lambda} {\hat S}^{\nu\lambda}
	- u^{\nu} n_{\lambda}  {\hat S}^{\mu\lambda} \, , \label{spin_def} \\
	(n_{\nu} + p_{\nu}/m) {\hat S}^{\mu\nu}=0\,,
\end{gather}
as generalizations of Eqs. (\ref{flat_center}), (\ref{flat_spin_def})
and (\ref{flat_SSC}) to curved spacetime. Note that $n_{\mu}$ now introduces the lapse function into these
expressions. Equation (\ref{ohashi}) was found to be the stress-energy tensor of a spinning
particle with mass dipole moment $m^{\mu}$ in \cite{O03}, i.e., its position variable is
the Newton-Wigner one in the Minkowski limit. Unfortunately, an explicit calculation shows
that the components $N(\sqrt{-g}T^{00})_{\rm shifted}$ and $g_{i\nu}(\sqrt{-g}T^{0\nu})_{\rm shifted}$
still depend on lapse and shift, which is not compatible with the ADM formalism. The solution
to this problem is inspired by the observation that multiplication with $n_{\mu}$ and
$g_{i\nu}$ does not commute with taking the Lie-derivative.

Therefore, we first calculate the projections of the stress-energy
tensor density with covariant SSC given by Eq.\ (\ref{stress_energy}),
i.e., $\sqrt{\gamma}T^{\mu\nu}n_{\mu}n_{\nu}$ and $- \sqrt{\gamma}T^{\nu}_i n_{\nu}$.
These quantities, after a long calculation, turn out to be independent of lapse and shift.
Adding up their Lie-shifted expressions (notice $m^0 = 0$) and also using the definitions (\ref{spin_def}) and
\be
\tilde{p}_i \equiv mu_i - n_{\mu}S^{k\mu}K_{ik} \, ,
\ee 
which fortunately eliminates $K_{ij}$ and therewith $\pi^{ij}$, we end up
with the expressions
\begin{align}
(\sqrt{\gamma}T^{\mu\nu}n_{\mu}n_{\nu})_{\rm shifted} &=
	- n\tilde{p}\delta  - \left(\gamma^{ij}\gamma^{kl}
		\frac{\tilde{p}_l}{m-n\tilde{p}}{\hat S}_{jk}\delta\right)_{,i}
	\equiv  N^2 \sqrt{\gamma} \tilde{T}^{00} \, , \\
\begin{split}
(- \sqrt{\gamma}T^{\nu}_i n_{\nu})_{\rm shifted} &=
	\tilde{p}_i\delta + \frac{1}{2} \left[\left(\gamma^{mk}{\hat S}_{ik}
	- (\gamma^{mk}\delta_i^p + \gamma^{mp} \delta_i^k)
		\gamma^{ql} {\hat S}_{qp}\frac{\tilde{p}_l\tilde{p}_k}{n\tilde{p} (m-n\tilde{p} )}
	\right)\delta\right]_{;m}\\
	&\quad + \delta x^l (\tilde{p}_{i;l} + \tilde{p}_{l,i} - \tilde{p}_{i,l})\delta
	\equiv  N \sqrt{\gamma} \tilde{T}^0_i \, ,
\end{split}
\end{align}
where $\delta x^l = - m^l/m$.
Full agreement is obtained with our previous results (\ref{H}) and (\ref{Hi})
if the linear momentum $\tilde{p}_i$ (as function of space and time coordinates)
gets parallel shifted along $\delta x^l$ and shows no rotation. Then
$p_i$ and $\tilde{p}_i$ play identical roles and may be identified and thus,
$\hat{T}^{\mu\nu}$ and $\tilde{T}^{\mu\nu}$ too. In order to fulfill the global
Poincar\'e algebra, we must indeed drop this term proportional to $\delta x^l$.
Including it into the definition of our canonical momentum (\ref{def_P}) is not
possible, see Sec. \ref{poincare}.

\section{Consistency considerations\label{consistency}}

Our action (\ref{action}) has the important properties that it exactly coincides
with the expected spin dynamics in the Minkowski case, that it reduces to the
usual point-mass dynamics for vanishing spins, and that our spin variables have 
constant Euclidean length like in the covariant equations of motion
approach for spin, see \cite{DJS07}. Our action, formally valid up to arbitrary order, thus
defines a spin dynamics that should at least be a good approximation to the
dynamics described by the covariant stress-energy tensor (\ref{stress_energy}).
We will argue in the following that up to the second post-Newtonian order, i.e., the
next-to-leading spin-orbit and spin(1)-spin(2) order, our dynamics is indeed
the same as of the covariant stress-energy tensor treated as source in
the Einstein field equations, see, e.g., \cite{FBB06}.

First we define
\be
	q^i \equiv - \gamma^{ij}\gamma^{kl}\frac{\hat{S}_{jk} P_l}{m-nP} \delta \, , \qquad
	r_i^k \equiv \frac{1}{2} \gamma^{km}{\hat S}_{im} \delta
		- \gamma^{ml}\gamma^{nk}\frac{\hat{S}_{l(i}P_{n)}P_m}{nP(m-nP)} \delta \, .
\ee
Then (\ref{Hcanon}) and (\ref{Hicanon}) are simply given by
$\mathcal{H}^{\rm matter} = - nP \delta - \tfrac{1}{2} t_{ij}^{k} \gamma^{ij}_{~~,k}
+ q^i_{, i}$ and
$\mathcal{H}_i^{\rm matter} = P_i\delta + r_{i,k}^k$.
Instead of (\ref{local_cons1}) we now have:
\begin{gather}
\frac{\delta {\cal{H}}^{\rm matter}(\mathbf{x})}{\delta \gamma^{ij}(\mathbf{x}')} =
	\frac{1}{2}\left[ - \frac{P_i P_j}{nP} \delta + t_{ij,k}^k(\mathbf{x})
	\right] \delta(\mathbf{x} - \mathbf{x}')
	- \frac{1}{2} \frac{\delta t_{kl}^m(\mathbf{x})}{\delta \gamma^{ij}(\mathbf{x}')}
		\gamma^{kl}_{~~,m}(\mathbf{x})
	+ \left[ \frac{\delta q^k(\mathbf{x})}{\delta \gamma^{ij}(\mathbf{x}')}
	- \frac{1}{2} t_{ij}^k(\mathbf{x}) \delta(\mathbf{x} - \mathbf{x}')
	\right]_{, k} \label{local_cons_violation1}\,, \\
\frac{\delta {\cal{H}}_k^{\rm matter}(\mathbf{x})}{\delta \gamma^{ij}(\mathbf{x}')} =
	\left[ \frac{\delta r_k^l(\mathbf{x})}{\delta \gamma^{ij}(\mathbf{x}')}
	\right]_{, l}\,. \label{local_cons_violation2}
\end{gather}
At the leading order the total divergences in (\ref{local_cons_violation1}) and
(\ref{local_cons_violation2}) do not contribute to (\ref{global_cons1}):
\be\label{approx_consist}
\frac{\delta H^{\rm matter}}{\delta \gamma^{ij}} =
	\frac{1}{2} N \sqrt{\gamma}\hat{T}_{ij} + \Order{(G)}\,.
\ee
Note that $\sqrt{\gamma}\hat{T}_{ij}$ is here a Minkowski expression, where our
variables are definitely the correct canonical ones. This ensures that the
evolution equations of $h^{\text{TT}}_{ij}$ and $\pi_{\text{TT}}^{ij}$ are correct at the
leading order, which is sufficient for a second post-Newtonian
Hamiltonian for spin, see also Eqs. (\ref{htt1}) and (\ref{htt2}).

The structure of (\ref{momentum}) and (\ref{amomentum}) is very promising, as it
already implies the fulfillment of a major part of the Poincar\'e algebra.
This is a very strong argument for our spin variables to be canonical up to
the second post-Newtonian order for spin and also in the spatial conformally flat case,
or, from a different point of view, for our spin dynamics (\ref{action}) to be
physical. This argument applies to the reduced phase space in the ADMTT
gauge, also recall $e_{ij}=e_{ji}$, where (\ref{momentum}) and
(\ref{amomentum}) were derived. The problems encountered for a gauge independent
formulation are briefly presented in Appendix \ref{curved_algebra}.

In addition, the next-to-leading order gravitational spin-orbit coupling
we will obtain in Sec.\ \ref{application} is the same as in
\cite{DJS07}. The latter was based on a completely different
approach using only the equations of motion for spin; the stress-energy
tensor for spin was not needed. Also the remaining generator $G^i$ of the Poincar\'e group was
determined there and the Poincar\'e invariance was shown for the two-body case.
In Sec.\ \ref{poincare} we will extend the proof of the Poincar\'e
invariance to the spin(1)-spin(2) interaction case.

Finally we present a nice property of our spin variable, both in the second
post-Newtonian approximation for spin and the spatial conformally flat case. In both cases
we can set $\hat{S}^{~~j}_{a i} \equiv \hat{S}_{ail}\gamma^{lj}
= \hat{S}_{alj}\gamma^{li} \equiv \hat{S}^i_{aj} = S_{a (i)(j)}$.
This is obvious in the spatial conformally flat case. The neglected $h^{\text{TT}}_{ij}$
contributions are merely total divergences at the second post-Newtonian order in the
Hamilton constraint, which do not contribute to the corresponding Hamiltonian.

\section{Applications\label{application}}

In this section we will derive within our formalism the ADM Hamiltonian
of two spinning compact bodies with next-to-leading order gravitational spin-orbit coupling,
recently obtained in \cite{DJS07}, and with next-to-leading order gravitational
spin(1)-spin(2) coupling. Some calculations
in this and the following Sections were confirmed with the help of xTensor \cite{MG}, a free package
for Mathematica \cite{W91}.

First we have to solve the constraints iteratively within the post-Newtonian
perturbation expansion, which can be seen as a formal expansion in $c^{-1}$.
In the source terms of the constraint equations, the action of the mass
$m$ has to be counted as $m \sim \Order(G c^{-2})$,  and similarly ${\bf P} \sim \Order(G c^{-3})$
and ${\bf S} \sim \Order(G c^{-3})$. In the following a subscript in round brackets
denotes the formal order in $c^{-1}$. We further set $\psi \equiv 1 + \phi/8$ and
$\pi^{ij}=\tilde{\pi}^{ij}+\pi_{\text{TT}}^{ij}$ in our coordinate conditions (\ref{cc}).
$\tilde{\pi}^{ij}$ can be written in terms of the vectors
$\tilde{\pi}^i \equiv \Delta^{-1} \pi^{ij}_{~~,j} = \Delta^{-1} \tilde{\pi}^{ij}_{~~,j}$ and
$\pi^i \equiv (\delta_{ij} - \frac{1}{2} \pa_i \pa_j \Delta^{-1}) \tilde{\pi}^j$ as
\begin{align}
	\tilde{\pi}^{ij} &= \pi^i_{, j} + \pi^j_{, i}
		- \delta_{ij} \pi^k_{, k} + \Delta^{-1} \pi^k_{,ijk} \\
                &= \tilde{\pi}^i_{, j} + \tilde{\pi}^j_{, i}
		- \tfrac{1}{2} \delta_{ij} \tilde{\pi}^k_{, k}
		- \tfrac{1}{2} \Delta^{-1} \tilde{\pi}^k_{, ijk} \, . \label{pidecomp}
\end{align}
The Hamilton constraint for an arbitrary source,
\be
	\frac{1}{16\pi\sqrt{\gamma}} \left[ \gamma \text{R}
		+ \frac{1}{2} \left( \gamma_{ij} \pi^{ij} \right)^2
		- \gamma_{ij} \gamma_{k l} \pi^{ik} \pi^{jl}\right]
		= {\cal{H}}^{\rm matter} ,
\ee
to the order needed for a second post-Newtonian Hamiltonian for spin, then reads
\begin{align}
	- \frac{1}{16\pi} \Delta \phi_{(2)} &= \mathcal{H}^{\rm matter}_{(2)}\,, \qquad
	- \frac{1}{16\pi} \Delta \phi_{(4)} = \mathcal{H}^{\rm matter}_{(4)}
		- \frac{1}{8} \mathcal{H}^{\rm matter}_{(2)} \phi_{(2)}\,, \\
	- \frac{1}{16\pi} \Delta \phi_{(6)} &= \mathcal{H}^{\rm matter}_{(6)}
		- \frac{1}{8} ( \mathcal{H}^{\rm matter}_{(4)} \phi_{(2)}
		+ \mathcal{H}^{\rm matter}_{(2)} \phi_{(4)} )
		+ \frac{1}{64} \mathcal{H}^{\rm matter}_{(2)} \phi_{(2)}^2
		+ \frac{1}{16\pi} \left[ ( \tilde{\pi}^{i j}_{(3)} )^2
		- \frac{1}{2} \phi_{(2) , i j} h^{\text{TT}}_{(4) i j} \right]
		\,, \label{phi6_const} \\
\begin{split}
	- \frac{1}{16\pi} \Delta \phi_{(8)} &= \mathcal{H}^{\rm matter}_{(8)}
		- \frac{1}{8} ( \mathcal{H}^{\rm matter}_{(6)} \phi_{(2)}
			+ \mathcal{H}^{\rm matter}_{(4)} \phi_{(4)}
			+ \mathcal{H}^{\rm matter}_{(2)} \phi_{(6)} )
		+ \frac{1}{64} ( \mathcal{H}^{\rm matter}_{(4)} \phi_{(2)}^2
		+ 2 \mathcal{H}^{\rm matter}_{(2)} \phi_{(2)} \phi_{(4)} ) \\
	&\quad	- \frac{1}{512} \mathcal{H}^{\rm matter}_{(2)} \phi_{(2)}^3
		+ \frac{1}{16\pi}\left[ \frac{1}{8} \phi_{(2)} (\tilde{\pi}^{i j}_{(3)})^2
		+ 2 \tilde{\pi}^{i j}_{(3)} \tilde{\pi}^{i j}_{(5)}
		- \frac{1}{16} \phi_{(2) , i} \phi_{(2) , j} h^{\text{TT}}_{(4) i j}
		+ \frac{1}{4} (h^{\text{TT}}_{(4) i j , k})^2 \right]
		+ (\text{td}) \,,
\end{split}
\end{align}
where (td) denotes total divergence, and the momentum constraint,
\be
	- \frac{1}{8\pi} \gamma_{ij} \pi^{jk}_{~~ ; k} = {\cal{H}}^{\rm matter}_i \, ,
\ee
can be expanded as
\begin{align}
	\frac{1}{16\pi}\tilde{\pi}^{i j}_{(3) , j} &=
		- \frac{1}{2} \mathcal{H}^{\rm matter}_{(3) i}\,, \\
	\frac{1}{16\pi}\tilde{\pi}^{i j}_{(5) , j} &=
		- \frac{1}{2} \mathcal{H}^{\rm matter}_{(5) i}
		- \frac{1}{32\pi} ( \phi_{(2)} \tilde{\pi}^{i j}_{(3)} )_{, j} \,.\label{pi5_const}
\end{align}
The solution to the partial differential equation
$\tilde{\pi}^{i j}_{~~ , j} = \Delta \tilde{\pi}^i = A^i$ for $\tilde{\pi}^{i j}$ is
given by (\ref{pidecomp}) and $\tilde{\pi}^i = \Delta^{-1} A^i$. The ADM Hamiltonian
can now be calculated via Eq.\ (\ref{conserved2}).

In the near-zone $h^{\text{TT}}_{(4) i j}$ results from:
\begin{align}
	\Delta h^{\text{TT}}_{(4) i j} &= \delta^{\text{TT} k l}_{i j}
		\left[ 32\pi \frac{\delta \left(
			\int{ d^3 x \, \mathcal{H}^{\rm matter}_{(8)}  } \right)}
			{\delta h^{\text{TT}}_{(4) k l}}
			- \frac{1}{4} \phi_{(2) , k} \phi_{(2) , l}
                        \right] \label{htt1}  \\
	&= \delta^{\text{TT} k l}_{i j} \left[- 16 \pi T_{(4) k l}
		- \frac{1}{4} \phi_{(2) , k} \phi_{(2) , l} \right]\,. \label{htt2}
\end{align}
The first of these equations is a consequence of the evolution equations (\ref{red_evol_eq}),
the second is a direct consequence of the Einstein equations. Both lead to the same result,
if the consistency condition (\ref{global_cons1}) is valid at the leading order. At this
order $\pi^{ij}_{\text{TT}}$ vanishes in the near-zone, the transition to the Routhian (\ref{routh})
is therefore trivial.

Now we introduce new indices $a$ and $b$ that number the spinning particles.
Expanding (\ref{Hcanon}) for a many-particle system yields
\begin{align}
	\mathcal{H}^{\rm matter}_{(2)} &= \sum_a m_a \delta_a\,, \qquad
	\mathcal{H}^{\rm matter}_{(4)} = \sum_a \left[ \frac{{\bf P}^2_a}{2 m_a} \delta_a
		+ \frac{1}{2 m_a} P_{a i} S_{a (i) (j)} \delta_{a , j} \right] \,,\\
	\mathcal{H}^{\rm matter}_{(6)} &= \sum_a \left[
		- \frac{({\bf P}^2_a)^2}{8 m_a^3} \delta_a
		- \frac{{\bf P}^2_a}{4 m_a} \phi_{(2)} \delta_a
		+ \frac{1}{4 m_a} P_{a i} S_{a (i) (j)} \phi_{(2) , j} \delta_a
		- \frac{{\bf P}^2_a}{8 m^3_a} P_{a i} S_{a (i) (j)} \delta_{a , j}
		- \frac{1}{4 m_a} P_{a i} S_{a (i) (j)} ( \phi_{(2)} \delta_a )_{, j}
	\right]\,, \label{Hm6} \\
	\mathcal{H}^{\rm matter}_{(8)} &= \sum_a \left[
		\frac{({\bf P}^2_a)^3}{16 m^5_a} \delta_a
		+ \frac{({\bf P}^2_a)^2}{8 m^3_a} \phi_{(2)} \delta_a
		+ \frac{5 {\bf P}^2_a}{64 m_a} \phi_{(2)}^2 \delta_a
		- \frac{{\bf P}^2_a}{4 m_a} \phi_{(4)} \delta_a
		- \frac{1}{2 m_a} P_{a i} P_{a j} h^{\text{TT}}_{(4) i j} \delta_a
		- \frac{{\bf P}_a^2}{8 m_a^3} P_{a i} S_{a (i) (j)} \phi_{(2) , j} \delta_a
	\right. \nonumber \\
	&\quad	\left. - \frac{5}{32 m_a} P_{a i} S_{a (i) (j)} \phi_{(2)} \phi_{(2) , j} \delta_a
		+ \frac{1}{4 m_a} P_{a i} S_{a (i) (j)} \phi_{(4) , j} \delta_a
		+ \frac{1}{2 m_a} P_{a i} S_{a (j) (k)} h^{\text{TT}}_{(4) i j , k} \delta_a
	\right] + (\text{td})\,, \label{H8}
\end{align}
and (\ref{Hicanon}) reads
\begin{align}
	\mathcal{H}^{\rm matter}_{(3) i} &= \sum_a \left[ P_{a i} \delta_a -
		\frac{1}{2} S_{a (j) (i)} \delta_{a , j} \right]\,, \\
	\mathcal{H}^{\rm matter}_{(5) i} &=
		\sum_a \frac{1}{4 m_a^2} ( P_{a i} P_{a j} S_{a (j) (k)} \delta_{a , k}
		+ P_{a j} P_{a k} S_{a (j) (i)} \delta_{a , k} )\,. \label{Hi5}
\end{align}
The leading order of (\ref{Tij}) is
\be
	T_{(4) i j} = \sum_a \frac{1}{2 m_a} ( 2 P_{a i} P_{a j} \delta_a
		+ P_{a i} S_{a (j) (k)} \delta_{a , k} + P_{a j} S_{a (i) (k)} \delta_{a , k} )\,.
\ee
The equivalence of (\ref{htt1}) and (\ref{htt2}) can now explicitly be checked.
The source terms of $\phi_{(4)}$, $\tilde{\pi}^{i j}_{(3)}$ and $h^{\text{TT}}_{(4) i j}$
arise from the point-mass source-terms by a substitution
$P_{a i} \rightarrow P_{a i} + \frac{1}{2} S_{(i) (j)} \pa_j$. As this
substitution commutes with $\Delta^{-1}$ and $\delta^{\text{TT} k l}_{i j}$, we can just
apply this substitution to the point-mass solutions of $\phi_{(4)}$,
$\tilde{\pi}^{i j}_{(3)}$ and $h^{\text{TT}}_{(4) i j}$, which are, e.g., in \cite{JS98}.
The results are, with $r_a = |{\bf x}-{\bf Z}_a|$,
\begin{align}
	\phi_{(4)}^{\text{spin}} &=
		2 \sum_a  \frac{P_{a i} S_{a (i) (j)}}{m_a} \left( \frac{1}{r_a} \right)_{, j}\,, \\
	\tilde{\pi}^{i j \, \text{spin}}_{(3)} &= - \sum_a
		\left[ S_{a (k) (i)} \left( \frac{1}{r_a} \right)_{, k j}
		+  S_{a (k) (j)} \left( \frac{1}{r_a} \right)_{, k i} \right]\,, \\
	h^{\text{TT} \, \text{spin}}_{(4) i j} &=
		\sum_a \frac{P_{a m} S_{a (k) (l)}}{m_a} \Bigg[
			( 4 \delta_{k ( i} \delta_{j ) m} \pa_l
			- 2 \delta_{i j} \delta_{k m} \pa_l ) \frac{1}{r_a}
		+ ( \delta_{k m} \pa_i \pa_j \pa_l
			- 2 \delta_{k ( i} \pa_{j )} \pa_m \pa_l ) r_a
		\Bigg]\,.
\end{align}
In order to get this expression for $h^{\text{TT}}_{(4) i j}$, it is actually easier
to solve (\ref{htt1}) directly, utilizing the formula $8 \pi \Delta^{-2} \delta_a = - r_a$,
than to use the substitution. The unknown functions $\phi_{(6)}$ and $\tilde{\pi}^{i j}_{(5)}$
are not needed for the second post-Newtonian Hamiltonian
$H_{\rm 2PN} = - \frac{1}{16 \pi} \int{ d^3 x \, \Delta \phi_{(8)} }$, they disappear
after some partial integrations. $\phi_{(6)}$ can be eliminated by
\be
	\int{ d^3 x \,\mathcal{H}^{\rm matter}_{(2)} \phi_{(6)} } =
		- \frac{1}{16\pi} \int{ d^3 x \, (\Delta \phi_{(2)}) \phi_{(6)} }
	= - \frac{1}{16\pi} \int{ d^3 x \, \phi_{(2)} (\Delta \phi_{(6)}) }
\ee
and then using the constraint (\ref{phi6_const}) for $\phi_{(6)}$.
Using (\ref{pidecomp}) for $\tilde{\pi}^{i j}_{(3)}$ and also (\ref{pi5_const})
and (\ref{Hi5}), we get,
\begin{equation}\label{trick2}
	\int{ d^3 x \, \tilde{\pi}^{i j}_{(3)} \tilde{\pi}^{i j}_{(5)} }
	= - \frac{1}{2} \int{ d^3 x \, \tilde{\pi}^{i j}_{(3)} \left[
		\phi_{(2)} \tilde{\pi}^{i j}_{(3)}
		+ \sum_a \frac{1}{2 m_a^2} P_{a i} P_{a k} S_{a (k) (j)} \delta_a \right] }\,.
\end{equation}

The $h^{\text{TT}}_{(4) i j}$ part of the Hamiltonian can also be simplified. We define $A_{(4) ij}$
such that $\Delta h^{\text{TT}}_{(4) i j} = \delta^{\text{TT} k l}_{i j} A_{(4) kl}$:
\be
	\frac{1}{16 \pi} A_{(4) ij} \equiv - \sum_a \frac{P_{a i} P_{a j}}{m_a} \delta_a
		- \sum_a \frac{1}{m_a} P_{a i} S_{a (j) (n)} \delta_{a , n}
		- \frac{1}{4} \phi_{(2) , i} \phi_{(2) , j}\,.
\ee
The $h^{\text{TT}}_{(4) i j}$ contribution to the Hamiltonian then is
\be
	\frac{1}{16 \pi} \int d^3 x \, \left[ \frac{1}{2} A_{(4) ij} h^{\text{TT}}_{(4) i j}
		+ \frac{1}{4} (h^{\text{TT}}_{(4) i j , k})^2 \right]
	= \frac{1}{16 \pi} \int d^3 x \, \frac{1}{4} A_{(4) ij} h^{\text{TT}}_{(4) i j} \, .
\ee
Here we used the fact that $\delta^{\text{TT} k l}_{i j}$ is a Hermitian operator,
$(h^{\text{TT}}_{(4) i j , k})^2 = - h^{\text{TT}}_{(4) i j} \Delta h^{\text{TT}}_{(4) i j} + (\text{td})$,
and of course $h^{\text{TT}}_{(4) i j} = \delta^{\text{TT} k l}_{i j} h^{\text{TT}}_{(4) kl}$.
The spin part of this can further be written as
\be\label{htt_trick}
	\frac{1}{16 \pi} \int d^3 x \, \left[
		\frac{1}{2} A_{(4) ij}^{\text{point-mass}}
			h^{\text{TT} \, \text{spin}}_{(4) i j}
		+ \frac{1}{4} A_{(4) ij}^{\text{spin}}
			h^{\text{TT} \, \text{spin}}_{(4) i j} \right] \, .
\ee
Note the factor $\frac{1}{2}$ instead of $\frac{1}{4}$ in the spin-orbit part.
This transformation of the $h^{\text{TT}}_{(4) i j}$ contribution is very convenient,
because the spin part of $h^{\text{TT}}_{(4) i j}$ is much simpler than its point-mass part,
which does not contribute any more in Eq.\ (\ref{htt_trick}).

The integral $H_{\rm 2PN} = - \frac{1}{16 \pi} \int{ d^3 x \, \Delta \phi_{(8)} }$
can now be computed. The regularization is, at the second post-Newtonian order,
done by Hadamard's partie finie method and by analytic regularization,
see, e.g., \cite{S85, J97, BF00}. In Appendix \ref{regularization} the formulas
needed to regularize the integrals occurring in this calculation are assembled.

\subsection{Results for $H_{\text{SO}}^{\text{NLO}}$  and $H_{\text{SS}}^{\text{NLO}}$}

Now we are ready to present the results. Our Hamiltonian for two spinning compact bodies
has a next-to-leading order spin-orbit part
$H_{\text{SO}}^{\text{NLO}}$ and a next-to-leading order
spin(1)-spin(2) part $H_{\text{SS}}^{\text{NLO}}$ given by:
\begin{align}
\begin{split}\label{HSO}
	H_{\text{SO}}^{\text{NLO}} &=
			- \frac{\picSin}{r_{1 2}^2}
			\left[ \frac{5 m_2 \pipi}{8 m_1^3} + \frac{3 \pipii}{4 m_1^2}
			- \frac{3 \piipii}{4 m_1 m_2}
			+ \frac{3 \pin \piin}{4 m_1^2}
			+ \frac{3 \piin^2 }{2 m_1 m_2} \right] \nl
		+ \frac{\piicSin}{r_{12}^2}
			\left[ \frac{\pipii}{m_1 m_2} + \frac{3 \pin \piin }{m_1 m_2} \right] \nl
		+ \frac{\picSipii}{r_{12}^2}
			\left[ \frac{2 \piin}{m_1 m_2} - \frac{3 \pin}{4 m_1^2} \right] \nl
		- \frac{\picSin}{r_{12}^3} \left[
			  \frac{11 m_2}{2} + \frac{5 m_2^2}{m_1}
		\right]
		+ \frac{\piicSin}{r_{1 2}^3} \left[
			  6 m_1 + \frac{15 m_2}{2}
		\right]
	+ (1 \leftrightarrow 2)\,,
\end{split}\\
\begin{split}\label{HSS}
	H_{\text{SS}}^{\text{NLO}} &= \frac{1}{2 m_1 m_2 r_{1 2}^3} [
			\tfrac{3}{2} \picSin \piicSiin
			+ 6 \piicSin \picSiin \nlq
			- 15 \Sin \Siin \pin \piin
			- 3 \Sin \Siin \pipii \nlq + 3 \Sipii \Siin \pin
			+ 3 \Siipi \Sin \piin + 3 \Sipi \Siin \piin \nlq
			+ 3 \Siipii \Sin \pin - \tfrac{1}{2} \Sipii \Siipi
			+ \Sipi \Siipii \nlq - 3 \SiSii \pin \piin + \tfrac{1}{2} \SiSii \pipii
		] \nl
		+ \frac{3}{2 m_1^2 r_{1 2}^3} [
			- \picSin \picSiin
			+ \SiSii \pin^2 - \Sin \Siipi \pin
		] \nl
		+ \frac{3}{2 m_2^2 r_{1 2}^3} [
			- \piicSiin \piicSin
			+ \SiSii \piin^2 - \Siin \Sipii \piin
		] \nl
		+ \frac{6 ( m_1 + m_2 )}{r_{1 2}^4} [ \SiSii - 2 \Sin \Siin ]\,.
\end{split}
\end{align}
Here $r_{12}= |{\bf Z}_1-{\bf Z}_2|$ is the euclidean distance between the two
particles and ${\bf n}_{12}$ denotes the unit vector $r_{12} {\bf n}_{12}= {\bf Z}_1 - {\bf Z}_2$.
$(1 \leftrightarrow 2)$ stands for repeating the preceding terms with particle one and two exchanged.
$H_{\text{SO}}^{\text{NLO}}$ is identical to the result in \cite{DJS07}.
The result for $H_{\text{SS}}^{\text{NLO}}$, already announced in \cite{SHS07},
differs from the corresponding spin(1)-spin(2) potential, 
$V^{\text{SS}}_{\text{3PN}}$, in \cite{PR06}. A canonical transformation connecting both results could
not be found \cite{SHS07}. In a recent preprint \cite{PR07}, prompted
by the preprint version of \cite{SHS07}, a missing contribution in Eq.\ (4) of \cite{PR06} has
been identified, see \cite{PR07}, [Eq.\ (2)], using information from
\cite{P07}, [Eq.\ (18)].

The term $-\frac{1}{16 \pi} \int d^3 x \, \phi_{(2)} ( \tilde{\pi}^{i j}_{(3)})^2$,
that contributes to the Hamiltonian via (\ref{trick2}), is the only one where terms
proportional to $\mathbf{S}_1^2$ and $\mathbf{S}_2^2$ survived the regularization procedure.
These terms must be dropped, because we already neglected them in the stress-energy tensor.

Of course we are also able to calculate the leading order spin-orbit and spin(1)-spin(2)
Hamiltonians via $H_{\rm 1PN} = - (16 \pi)^{-1} \int{ d^3 x \, \Delta \phi_{(6)} }$,
which gives the well-known results:
\begin{align}
	H_{\text{SO}}^{\text{LO}} &=
 		\sum_a \sum_{b \neq a}
		\frac{S_{a (i) (j)}}{r_{a b}^2} \left[ \frac{3 m_b}{2 m_a} n_{a b}^i p_{a j}
		- 2 n_{a b}^i p_{b j} \right] \,, \label{HSOlead} \\
	H_{\text{SS}}^{\text{LO}} &=
		\frac{1}{2} \sum_a \sum_{b \neq a}
			\frac{S_{a (k) (i)} S_{b (k) (j)}}{r_{a b}^3} [ \delta_{i j}
			- 3 n_{a b}^i n_{a b}^j ]\,.
\end{align}
Here $r_{ab}= |{\bf Z}_a-{\bf Z}_b|$ and $r_{ab} n_{ab}^i = \hat{z}^i_a - \hat{z}^i_b$.
These formulas are even valid for arbitrary many particles.

\section{Different derivation of $H_{\text{SS}}^{\text{NLO}}$ \label{treatment}}

In order to confirm our result for $H_{\text{SS}}^{\text{NLO}}$,
we use the method from \cite{DJS07} to rederive
$H_{\text{SS}}^{\text{NLO}}$. Our ansatz for
$H_{\text{SS}}^{\text{NLO}}$ linear in
${\bf S}_1$ and ${\bf S}_2$ is now:
\be\label{HSScheck}
	H_{\text{SS}}^{\text{NLO}} =
		\tilde{\Omega}_{(4) i j} S_1^{(i)} S_2^{(j)}
	= \mathbf{\Omega}_{(4)}^{\text{spin(2)}} \cdot \mathbf{S}_1
		= \mathbf{\Omega}_{(4)}^{\text{spin(1)}} \cdot \mathbf{S}_2\,.
\ee
Note that the equal signs are correct here, because
\be
	\left[ \mathbf{\Omega}_{(4)}^{\text{spin(2)}} \right]_i
		=  \tilde{\Omega}_{(4) i j} S_2^{(j)}
\ee
already includes the full dependence of the Hamiltonian on ${\bf S}_2$.
The formula for $\mathbf{\Omega}_{(4)}$ given in \cite{DJS07} can be used
without further changes, but now the spin-dependent parts of the quantities
have to be inserted. The evolution equations, correctly given by (\ref{evol_eq})
if (\ref{global_cons2}) and (\ref{global_cons1}) are fulfilled, read:
\begin{align}
	\gamma_{ij , 0} &= 2 N \gamma^{-1/2}
		(\pi_{ij} - \tfrac{1}{2} \gamma_{ij} \gamma_{kl} \pi^{kl})
		+ N_{i ; j} + N_{j ; i}\,, \label{detshift} \\
\begin{split}\label{detlapse}
	\pi^{ij}_{~~ , 0} &= - N \sqrt{\gamma} (\text{R}^{ij} - \tfrac{1}{2} \gamma^{ij} \text{R})
		+ \tfrac{1}{2} N \gamma^{-1/2} \gamma^{ij} (\pi^{mn} \pi_{mn}
		- \tfrac{1}{2} (\gamma_{mn} \pi^{mn})^2) \\
	&\quad	- 2 N \gamma^{-1/2} (\gamma_{mn} \pi^{im} \pi^{nj}
		- \tfrac{1}{2} \gamma_{mn} \pi^{mn} \pi^{ij})
		+ \sqrt{\gamma} (N^{; ij} - \gamma^{ij} N^{;m}_{~~~ ;m} ) \\
	&\quad	+ (\pi^{ij} N^m)_{;m} - N^i_{~;m} \pi^{mj} - N^j_{~ ;m} \pi^{mi}
		+ 8 \pi N \gamma^{im} \gamma^{jn} \sqrt{\gamma} T_{mn}\,.
\end{split}
\end{align}
Here $\text{R}^{ij}$ is the 3-dim.\ Ricci-tensor. Now we determine lapse and shift by demanding
that our coordinate conditions (\ref{cc}) are preserved under this time evolution.
In particular we insert (\ref{detshift}) into $3\gamma_{ij, 0 j} - \gamma_{jj, 0 i} = 0$,
and we take the $\delta_{ij}$-trace of (\ref{detlapse}). The post-Newtonian expansion
of the resulting expressions, with further simplifications using the constraints, leads to:
\begin{align}
	N_{(0)} &= 1 \,, \qquad
	N_{(2)} = - \frac{1}{4} \phi_{(2)} \,, \\
	\Delta N_{(4)} &= 4\pi T_{(4) i i} + 4\pi \mathcal{H}^{\rm matter}_{(4)}
			- \pi \mathcal{H}^{\rm matter}_{(2)} \phi_{(2)}
			+ \frac{1}{16} ( \phi_{(2)} \phi_{(2) , i} )_{,i} \,, \\
	\Delta N_{(3) i} + \frac{1}{3} N_{(3) j , j i} &=
		16\pi \mathcal{H}^{\rm matter}_{(3) i}\,, \\
	\Delta N_{(5) i} + \frac{1}{3} N_{(5) j , j i} &=
		16\pi \mathcal{H}^{\rm matter}_{(5) i}
		+ \left[ \phi_{(2)} \tilde{\pi}^{i j}_{(3)}
		+ N_{(3) ( j} \phi_{(2) , i )} \right]_{, j}
		- \frac{1}{3} [ N_{(3) j} \phi_{(2) , j} ]_{, i}\,.
\end{align}
Note that also $T_{ij}$ is needed for $N$. The solution of
$\Delta N_i + \frac{1}{3} N_{j , j i} = A_i$ is given by
$N_i = (\delta_{ij} - \frac{1}{4} \pa_i \pa_j \Delta^{-1}) \Delta^{-1} A_j$.
Again we can get $N_{(4)}$ and $N_{(3) i}$ by the substitution
$P_{a i} \rightarrow P_{a i} + \frac{1}{2} S_{(i) (j)} \pa_j$
from their point-mass solutions. This gives:
\be
	N_{(4)}^{\text{spin}} =
		- \frac{3}{2} \sum_a \frac{P_{a i} S_{a (i) (j)}}{m_a}
			\left( \frac{1}{r_a} \right)_{, j}\,, \qquad
	N_{(3) i}^{\text{spin}} =
		2 \sum_a S_{a (j) (i)} \left( \frac{1}{r_a} \right)_{, j}\,.
\ee
$N_{(5) i}$ is more complicated, but for $\mathbf{\Omega}_{(4)}$ we only need
\be
\begin{split}
	\epsilon_{i j k} N_{(5) j , k}^{\text{spin}}
	&= \epsilon_{i j k} \sum_a \left[
		- 2 \frac{P_{a s} P_{a m} S_{a (m) (t)}}{m^2_a}
			\delta_{j ( s} \delta_{t ) l} \left( \frac{1}{r_a} \right)_{, k l}
		+ m_a S_{a (j) (m)} \left( \frac{1}{r_a^2} \right)_{, k m}
		\right] \\
	&\quad	+ \epsilon_{i j k} \pa_k \pa_l \sum_a \sum_{b \neq a} \pa^a_m [
		4  m_b S_{a (m) (j)} ( \pa^b_l - \pa^a_l )
		+ 4 m_b S_{a (m) (l)} ( \pa^b_j - \pa^a_j )] \ln s_{a b} \, ,
\end{split}
\ee
where $s_{ab} = r_a + r_b + r_{ab}$ and $\pa^a_i$ and $\pa^b_i$
are partial derivatives with respect to ${\bf Z}_a$ and ${\bf Z}_b$,
and we used the formula $\Delta \ln s_{ab} = (r_a r_b)^{-1}$.
Finally, we get from the leading order spin-orbit Hamiltonian (\ref{HSOlead}):
\be
	v^{i \, \text{spin}}_{(3) a}
		= \{ \hat{z}^i_a , H_{\text{SO}}^{\text{LO}} \}
		= - \sum_{b \neq a} \left( \frac{3 m_b S_{a (i) (j)}}{2 m_a}
			+ 2 S_{b (i) (j)} \right) \frac{n_{a b}^j}{r_{a b}^2}\,.
\ee
Now $\mathbf{\Omega}_{(4) 1}^{\text{spin}}$ can be calculated
by applying partie finie regularization, e.g.,
\be
\begin{split}
	\frac{1}{2} S_{a (j) (k)} \text{Reg}_a
		\left( N_{(5) j , k}^{\text{spin}} \right)
	&= \frac{3}{2} \frac{P_{b i} P_{b m} S_{a (j) (k)} S_{b (n) (l)}}{m_b^2 r_{a b}^3}
			[ - \delta_{i j} \delta_{m n} n_{ab}^k n_{a b}^l
			+ \delta_{j n} \delta_{m l} n_{ab}^k n_{a b}^i ] \\
	&\quad	+ \frac{S_{a (i) (j)} S_{b (i) (l)}}{r_{a b}^4} ( 3 m_a + m_b ) ( 4 n_{ab}^j n_{a b}^l
			- \delta_{j l} ) \, , \label{pf_example}
\end{split}
\ee
where $a=1$ and $b=2$, or $a=2$ and $b=1$, and
$\text{Reg}_a(f(\mathbf{x}))=f_{\text{reg}}(\mathbf{Z}_a)$,
see Appendix \ref{regularization}. Although this term is not symmetric under
exchange of both particles, the final result (\ref{HSScheck}) recovers this
symmetry, and indeed turns out to be the same as (\ref{HSS}). It should be
stressed that this approach is indeed independent from the one of the last section,
in particular, lapse and shift functions had to be determined, also using $T_{ij}$,
and $\mathbf{\Omega}_{(4)}$ was determined using the equations of motion
of a spinning body in \cite{DJS07}.

\section{Approximate Poincar\'e algebra\label{poincare}}

At last, the Poincar\'e invariance at the next-to-leading spin(1)-spin(2)
order was not yet verified. First we calculate
${\bf G}_{\text{SO}}^{\text{NLO}}$ and
${\bf G}_{\text{SS}}^{\text{NLO}}$
with the help of (\ref{conserved2}),
i.e., ${\bf G}_{\rm 2PN} = - \frac{1}{16 \pi} \int{ d^3 x \, {\bf x} \Delta \phi_{(6)} }$.
Using the 3-particle integrals from Ref.\ \cite{JS97}
\begin{align}
	\int d^3 x \, \frac{r_{a}^2}{r_{b}r_{c}}
	&=-4\pi\left[\Delta^{-1}\frac{r_{a}^2}{r_{b}}\right]_{\mathbf{x}=\mathbf{Z}_c}
	=-4\pi\left[-\frac{1}{6}r_{bc}^{3}+\frac{1}{4}(r_{ac}^2+r_{ab}^2)r_{bc}\right]\,, \\
	\int d^3 x \, \frac{r_{a}^2r_{b}}{r_{c}}
	&=-4\pi\left[\Delta^{-1}(r_{a}^2r_{b})\right]_{\mathbf{x}=\mathbf{Z}_c}
	=-\frac{4\pi}{180}\left[10r_{ac}^2+5r_{ab}^2-4r_{bc}^2\right]r_{bc}^3 \,,
\end{align}
and treating the origin as a particle coordinate, results in
\begin{align}
\begin{split}
	\mathbf{G}_{\text{SO}}^{\text{NLO}} &=
		- \sum_a \frac{\mathbf{P}_a^2}{8 m_a^3} (\mathbf{P}_a \times \mathbf{S}_a)
		+ \sum_a \sum_{b \neq a} \frac{m_b}{4 m_a r_{ab}} \left[
			- 5 (\mathbf{P}_a \times \mathbf{S}_a)
			+ ((\mathbf{P}_a \times \mathbf{S}_a) \cdot \mathbf{n}_{ab})
				\frac{5\mathbf{Z}_a+\mathbf{Z}_b}{r_{ab}}
		\right] \\
	&\quad	+ \sum_a \sum_{b \neq a} \frac{1}{r_{ab}} \left[
			\frac{3}{2} (\mathbf{P}_b \times \mathbf{S}_a)
			- \frac{1}{2} (\mathbf{n}_{ab} \times \mathbf{S}_a)
				(\mathbf{P}_b \cdot \mathbf{n}_{ab})
			- ((\mathbf{P}_b \times \mathbf{S}_a) \cdot \mathbf{n}_{ab})
				\frac{\mathbf{Z}_a+\mathbf{Z}_b}{r_{ab}}
		\right] \, ,
\end{split}\\
	\mathbf{G}_{\text{SS}}^{\text{NLO}} &=
	\frac{1}{2} \sum_a \sum_{b \neq a} \left[(\mathbf{S}_{b}\cdot\mathbf{n}_{ab})
		\frac{\mathbf{S}_{a}}{r_{ab}^2}
	+\left(3(\mathbf{S}_{a}\cdot\mathbf{n}_{ab})(\mathbf{S}_{b}\cdot\mathbf{n}_{ab})
		-(\mathbf{S}_{a}\cdot\mathbf{S}_{b})\right)
		\frac{\mathbf{Z}_{a}}{r_{ab}^3}\right] \, .
\end{align}
We get the same result for
${\bf G}_{\text{SO}}^{\text{NLO}}$ as in \cite{DJS07}, if we consider our expression
for two particles.
Now the Poincar\'e algebra for two bodies, including
$H_{\text{SS}}^{\text{NLO}}$,
can be verified in the same way as in
\cite{DJS07}, also see \cite{DJS00}. The algebra is indeed fulfilled.

If we would have kept the term proportional to $\delta x^l$ in Sec.\ \ref{curved},
then we must include it into the definition of our canonical momentum (\ref{def_P}).
This gives only a change in $\mathcal{H}^{\rm matter}$, in particular, the term
$- \frac{{\bf P}_a^2}{8 m_a^3} P_{a i} S_{a (i) (j)} \phi_{(2) , j} \delta_a$
in Eq.\ (\ref{H8}) disappears. Then we must add
$\frac{\picSin}{r_{1 2}^2}\,\frac{m_2 \pipi}{2 m_1^3}+ (1 \leftrightarrow 2)$
to the Hamiltonian in Eq.\ (\ref{HSO}), but the center-of-mass vector $\mathbf{G}$,
calculated in this section, stays unchanged, as Eq.\ (\ref{H8}) does not contribute to it.
The Poincar\'e algebra would not be fulfilled any more, therefore we have to
drop the term proportional to $\delta x^l$ in Sec.\ \ref{curved}.

\acknowledgments
GS is grateful to Professor S. Deser for helpful discussions.
This work is supported by the Deutsche Forschungsgemeinschaft (DFG) through
SFB/TR7 ``Gravitational Wave Astronomy''.

\appendix
\section{Poisson-bracket algebra of spinning particles stress-energy tensor in Minkowski space\label{flat_algebra}}

The following equal-time algebra, i.e., $x^0 = x'^0$, must be valid in
Minkowski space \cite{BD67} (see also \cite{S98}),
\begin{align}
	\{ \mathcal{H}^{\rm matter}(x) , \mathcal{H}^{\rm matter}(x') \} =&
		- \left[ \mathcal{H}^{\rm matter}_i(x)
		+ \mathcal{H}^{\rm matter}_i(x') \right] \delta_{\mathbf{x} \mathbf{x}' , i}
		+ \pa^{~}_m \pa^{~}_n \pa'_p \pa'_q
			\left[ f_{m n p q}(x) \, \delta_{\mathbf{x} \mathbf{x}'} \right]\,, \\
	\{ \mathcal{H}^{\rm matter}_i(x) , \mathcal{H}^{\rm matter}(x') \} =&
		- \mathcal{H}^{\rm matter}(x) \, \delta_{\mathbf{x} \mathbf{x}' , i}
		- T_{i j}(x') \, \delta_{\mathbf{x} \mathbf{x}' , j}
		+ \pa^{~}_n \pa'_p \pa'_q
			\left[ g_{i n p q}(x) \, \delta_{\mathbf{x} \mathbf{x}'} \right]\,, \\
	\{ \mathcal{H}^{\rm matter}_i(x) , \mathcal{H}^{\rm matter}_j(x') \} =&
		- \mathcal{H}^{\rm matter}_j(x) \, \delta_{\mathbf{x} \mathbf{x}' , i}
		- \mathcal{H}^{\rm matter}_i(x') \, \delta_{\mathbf{x} \mathbf{x}' , j}
		+ \pa^{~}_n \pa'_q
			\left[ h_{i n j q}(x) \, \delta_{\mathbf{x} \mathbf{x}'} \right]\,.
\end{align}
Here $\delta_{\mathbf{x} \mathbf{x}'} \equiv \delta(\mathbf{x} - \mathbf{x}')$,
where $\mathbf{x}$ and $\mathbf{x}'$ are the spatial parts of $x$ and $x'$.
This local algebra is a consequence of the global Poincar\'e algebra, whose generators
can be written in terms of integrals over certain components of the stress-energy tensor,
similar to Eq.\ (\ref{conserved}). The terms containing $f$, $g$ and $h$
turn into vanishing surface terms in the generators of the Poincar\'e algebra. It holds:
\begin{alignat}{2}
	f_{m n p q} &= f_{n m p q} = f_{m n q p},& \quad & \text{the same for $g$ and $h$}\,, \\
	f_{m n p q} &= - f_{p q m n},& \quad & \text{the same for $h$}\,.
\end{alignat}
An explicit calculation with the Minkowski versions of (\ref{Hcanon}), (\ref{Hicanon}) and (\ref{Tij})
shows that we have to set
\begin{gather}
	f_{m n p q}(x) = 0 = g_{i n p q}(x)\,, \\
	h_{i n j q}(x) = \bigg[ - \hat{S}_{q )( n} \mathcal{P}_{i )( j}
		- \delta^{k l} \frac{p_k \hat{S}_{l ( n} \mathcal{P}_{i )( j} p_{q )}}{(n p) (m - n p)}
		+ \delta^{k l} \frac{p_k \hat{S}_{l ( q} \mathcal{P}_{j )( i} p_{n )}}{(n p) (m - n p)} \bigg] \delta\,,
	\quad \text{with $\mathcal{P}_{i j} \equiv \delta_{i j} - \dfrac{p_i p_j}{(n p)^2}$} \,.\label{h_def}
\end{gather}
Now the local algebra is fulfilled linear in the spin variables, as it should be, because
our variables are known to be canonical in the Minkowski case. It was already noted in
\cite{S62}, in the context of quantum field theory, that $h_{i n j q}(x)$ does not
generally vanish if fields with spin are present, in particular, spin-$\frac{1}{2}$ fields. For fields
with spin $\frac{3}{2}$ one even has $f_{m n p q} \neq 0$, see \cite{S63b}.
The consequences of nonvanishing $h_{i n j q}(x)$ are considered
in Appendix B.

\section{Poisson-bracket algebra of non-spinning particles stress-energy tensor in general relativity\label{curved_algebra}}

We assume that the following equal-time constraint algebra on the
nonreduced phase space without gauge fixing is valid \cite{S63c,DW67,HRT76}:
\begin{align}
	\{ \mathcal{H}(x) , \mathcal{H}(x') \} =&
		- \left[ \mathcal{H}_i(x) \gamma^{i j}(x)
		+ \mathcal{H}_i(x') \gamma^{i j}(x') \right] \delta_{\mathbf{x} \mathbf{x}' , j}\,,
			\label{diffeo1} \\
	\{ \mathcal{H}_i(x) , \mathcal{H}(x') \} =&
		- \mathcal{H}(x) \, \delta_{\mathbf{x} \mathbf{x}' , i} \label{diffeo2}\,, \\
	\{ \mathcal{H}_i(x) , \mathcal{H}_j(x') \} =&
		- \mathcal{H}_j(x) \, \delta_{\mathbf{x} \mathbf{x}' , i}
		- \mathcal{H}_i(x') \, \delta_{\mathbf{x} \mathbf{x}' , j}\,. \label{diffeo3}
\end{align}
Note that, compared to the algebra of the last section, the $T_{i j}$ term and
the surface terms are absent. If this local algebra is fulfilled,
then the global Poincar\'e algebra can also be derived \cite{S63c, HRT76}. At this
point the constraints are not solved and no coordinate conditions are imposed,
i.e., one has to use
$\{\gamma_{ij}({\bf x},t), \pi^{kl}({\bf x}',t)\} = 16\pi \delta^{kl}_{ij}\delta({\bf x} - {\bf x}')$,
where $\delta^{kl}_{ij}=\delta^k_{(i}\delta^l_{j)}$.

Remember that ${\cal{H}}$ and ${\cal{H}}_i$ are a sum of matter and field parts.
The field-field Poisson-brackets cancel with the field terms
on the right hand side of each relation of the algebra, because the algebra is
fulfilled if no matter would be present, see \cite{DW67}.
In the context of (\ref{def_pi}), the matter parts do not depend on $\pi^{ij}$,
Eq.\ (\ref{local_cons2}). Therefore all terms proportional to $\pi^{ij}$ that
arise from the mixed matter-field Poisson-brackets must vanish separately.
These are exactly the ones with $\mathcal{H}^{\rm field}$:
\begin{gather}
	\{ \mathcal{H}^{\rm field}(x) , \mathcal{H}^{\rm matter}(x') \}
	+ \{ \mathcal{H}^{\rm matter}(x) , \mathcal{H}^{\rm field}(x') \} = 0 \,,\label{mixed1} \\
	\{ \mathcal{H}^{\rm matter}_i(x) , \mathcal{H}^{\rm field}(x') \} = 0\,.\label{mixed2}
\end{gather}
These conditions are indeed equivalent to Eq.\ (\ref{local_cons1}). From (\ref{mixed2}) follows:
\be\label{pi_prop_consequence}
	\pi^{kl}(\mathbf{x}') \left( \gamma_{kl}(\mathbf{x}') \gamma^{mn}(\mathbf{x}') \frac{\delta \mathcal{H}^{\rm matter}_i(\mathbf{x})}{\delta \gamma^{mn}(\mathbf{x}')}
	- 2 \frac{\delta \mathcal{H}^{\rm matter}_i(\mathbf{x})}{\delta \gamma^{kl}(\mathbf{x}')} \right) = 0
\ee
Because of the factor $2$, we have
$\frac{\delta \mathcal{H}^{\rm matter}_i(\mathbf{x})}{\delta \gamma^{kl}(\mathbf{x}')} = 0$
as the only possible solution. In Eq.\ (\ref{mixed1}) we make a general ansatz
\begin{align}
\frac{\delta {\cal{H}}^{\rm matter}(\mathbf{x})}{\delta
\gamma^{ij}(\mathbf{x}')} & =
	a_{ij}(\mathbf{x}) \delta(\mathbf{x} - \mathbf{x}')
	+ \sum_{n=1}^{N}
		a_{ij}^{~~ k_1 \ldots k_n}(\mathbf{x}) \partial_{k_1} \ldots \partial_{k_n} \delta(\mathbf{x} - \mathbf{x}')\,, \\
	a_{ij}^{~~ k_1 \ldots k_n}(\mathbf{x}) & \equiv
		\frac{\partial {\cal{H}}^{\rm matter}\left[ \gamma^{ij}, (\partial_{k_1} \gamma^{ij}), \ldots,
			(\partial_{k_1} \ldots \partial_{k_N} \gamma^{ij}) \right]}
		{\partial (\partial_{k_1} \ldots \partial_{k_n} \gamma^{ij})}(\mathbf{x}) \,,
\end{align}
integrate over $\mathbf{x}'$ and demand that the term with the highest number of
derivatives on $\pi^{ij}$ must vanish separately. The resulting equation is similar
to (\ref{pi_prop_consequence}), this time leading to $a_{ij}^{~~ k_1 ... k_N}(\mathbf{x})=0$.
Now we have effectively reduced $N$ by one, proceeding this way we get
$a_{ij}^{~~ k_1 ... k_n}(\mathbf{x})=0$ for all $n$ with $ 1 \leq n \leq N$. Comparing with (\ref{global_cons1})
we see that $a_{ij}$ has to be identified as $\frac{1}{2}\sqrt{\gamma} T_{ij}$.

From (\ref{local_cons1}) then immediately follows:
\begin{gather}
	\{ \mathcal{H}^{\rm field}_i(x) , \mathcal{H}^{\rm matter}(x') \} =
	\sqrt{\gamma} T_{jk}(x') \left[
			\delta^j_i \gamma^{k l}(x') \, \delta_{\mathbf{x} \mathbf{x}' , l}
		 	+ \gamma^{j k}_{~ ~ , i}(x') \, \delta_{\mathbf{x} \mathbf{x}'} \right]\,, \\
	\{ \mathcal{H}^{\rm field}_i(x) , \mathcal{H}_j^{\rm matter}(x') \} = 0\,.
\end{gather}
For the matter part therefore a similar algebra as in the Minkowski case has to hold,
\begin{align}
	\{ \mathcal{H}^{\rm matter}(x) , \mathcal{H}^{\rm matter}(x') \} =&
		- \left[ \mathcal{H}^{\rm matter}_i(x) \gamma^{i j}(x)
		+ \mathcal{H}^{\rm matter}_i(x') \gamma^{i j}(x') \right]
			\delta_{\mathbf{x} \mathbf{x}' , j} \,, \label{matter_algebra_1} \\
	\{ \mathcal{H}^{\rm matter}_i(x) , \mathcal{H}^{\rm matter}(x') \} =&
		- \mathcal{H}^{\rm matter}(x) \, \delta_{\mathbf{x} \mathbf{x}' , i}
		- \sqrt{\gamma} T_{jk}(x') \left[
			\delta^j_i \gamma^{k l}(x') \, \delta_{\mathbf{x} \mathbf{x}' , l}
		 	+ \gamma^{j k}_{~ ~ , i}(x') \, \delta_{\mathbf{x} \mathbf{x}'} \right]\,,
		 \label{matter_algebra_2} \\
	\{ \mathcal{H}^{\rm matter}_i(x) , \mathcal{H}^{\rm matter}_j(x') \} =&
		- \mathcal{H}^{\rm matter}_j(x) \, \delta_{\mathbf{x} \mathbf{x}' , i}
		- \mathcal{H}^{\rm matter}_i(x') \, \delta_{\mathbf{x} \mathbf{x}' , j}\,.
			\label{matter_algebra_3}
\end{align}
If the consistency conditions (\ref{local_cons2})
and (\ref{local_cons1}) hold, still in the context of (\ref{def_pi}), then this algebra can be used to validate the canonical
variables of the matter part on the nonreduced phase space.
This algebra is indeed fulfilled for point-masses, and of course also
Eqs.\ (\ref{def_pi}), (\ref{local_cons2}) and (\ref{local_cons1}).

Because in the algebra (\ref{matter_algebra_1}\nobreakdash--\ref{matter_algebra_2}) there are
no variations with respect to $\gamma_{ij}$ and $\pi^{ij}$ left any more, we can now
consider its Minkowski space limit. This gives the algebra of the last Section with
$0=f_{m n p q}(x)=g_{i n p q}(x)=h_{i n j q}(x)$. But in the last Section we have
seen that for spinning bodies $h_{i n j q}(x)$ does not vanish, not even in the
Newtonian case. Therefore we already see by inspecting the Minkowski case that
the coupling to gravity can not be of the simple kind defined by (\ref{def_pi}),
(\ref{local_cons2}) and (\ref{local_cons1}). Another problem that must be
addressed by a gauge independent formulation is that we have
(\ref{local_cons_violation1}) and (\ref{local_cons_violation2}) instead of
Eq.\ (\ref{local_cons1}). The total divergence in (\ref{local_cons_violation1})
contributes to the leading order. Therefore (\ref{local_cons1}) is not even
fulfilled at the leading order. This together with the nonvanishing
$h_{i n j q}(x)$ leads to additional contributions to the algebra
(\ref{diffeo1}\nobreakdash--\ref{diffeo3}) and suggests
its extension when spinning objects are present. Extensions to (\ref{def_pi})
may also be considered, recall Ref. \cite{K77}.

Important about the algebra of (first-class) constraints is their connection
to the gauge structure of the considered theory, see, e.g., \cite{C82}.
This makes the algebra (\ref{diffeo1}\nobreakdash--\ref{diffeo3}) quite robust
even if other systems are coupled to gravity. Extended forms of the algebra
(\ref{diffeo1}\nobreakdash--\ref{diffeo3}) were, to the best of our knowledge,
indeed only found when the gauge structure was also extended
\cite{T80,NT78,H83,CNP82}. This we will keep in mind when investigating
higher approximations in future.

The considerations of this Appendix do not show up any inconsistencies
of the canonical formulation for spinning bodies given in this paper because
they are requiring that no coordinate conditions are imposed.
Rather they indicate that the gauge conditions (\ref{cc})
must be seen as essential part of our formulation.

\section{Partie Finie and analytic regularization\label{regularization}}

In this Appendix we will present the regularization techniques we used in this work.
We first give a short overview of Hadamard's partie finie method. Let us consider
$f$ being a real function defined in an environment of the point
$\mathbf{x}_{0}\in\mathbb{R}^3$, except in this point where $f$ is singular.
We define a family of complex-valued functions $f_{\mathbf{n}}$ as follows:
\begin{equation}
f_{\mathbf{n}}\;:\;\mathbb{C}\ni\varepsilon\;\mapsto\;f_{\mathbf{n}}(\varepsilon)
\equiv f(\mathbf{x}_{0}+\varepsilon\mathbf{n})\;\in\mathbb{C}\,.
\end{equation}
We expand $f_{\mathbf{n}}$ in a Laurent-series around $\varepsilon=0$:
\begin{equation}\label{entop}
f_{\mathbf{n}}(\varepsilon)=\sum^{\infty}_{m=-N} a_{m}(\mathbf{n})\varepsilon^{m} \, .
\end{equation}
The regularized value of $f$ at $\mathbf{x}_{0}$ is defined as the
coefficient at $\varepsilon^{0}$ in the expansion (\ref{entop})
mean-valued over all unit vectors $\mathbf{n}$, \cite{S85, J97, BF00},
\begin{equation}
f_{\text{reg}}(\mathbf{x}_{0}) \equiv \frac{1}{4\pi}\oint\text{d}\Omega\,a_{0}(\mathbf{n})\,.
\end{equation}
This formula can be used to calculate integrals with delta-distributions. We define
\begin{equation}
\int d^3 x\,f(\mathbf{x})\delta(\mathbf{x}-\mathbf{Z}_{a}):=
f_{\text{reg}}(\mathbf{Z}_{a})\,,
\end{equation}
which provides us with a formula for calculating Poisson integrals of the form
\begin{equation}
\Delta^{-1}\left\lbrace\sum_{a}f(\mathbf{x})\delta_{a}\right\rbrace=
\Delta^{-1}\left\lbrace\sum_{a}f_{\text{reg}}(\mathbf{x})\delta_{a}\right\rbrace=
\sum_{a}f_{\text{reg}}(\mathbf{Z}_{a})\Delta^{-1}\delta_{a}=
-\frac{1}{4\pi}\sum_{a}f_{\text{reg}}(\mathbf{Z}_{a})\frac{1}{r_{a}}\,.
\end{equation}
A complicated example is given by (\ref{pf_example}).

Integrals that do not contain a delta function are regularized analytically \cite{J97}.
First we perform all differentiations in the integrand, and then constrain ourselves to the
two particle case. The integrand then depends on $r_1 = |{\bf x}-{\bf Z}_1|$,
${\bf n}_1 = ({\bf x}-{\bf Z}_1) / r_1$ and $r_2$, ${\bf n}_2$. Now we introduce the
analytic regularization parameter $\epsilon$ by replacing $r_1^{\alpha}$ by
$r_1^{\alpha + \mu \epsilon}$, and $r_2^{\beta}$ by $r_2^{\beta + \nu \epsilon}$.
The vectors $n_1^i$ and $n_2^j$ are then rewritten as partial
derivatives $\pa^1_i $ and $\pa^2_j$ with respect to the particle positions
\begin{gather}
	r_a^{\alpha} n_a^i = - \frac{\pa^a_i r_a^{\alpha+1}}{\alpha + 1}\,, \qquad
	r_a^{\alpha} n_a^i n_a^j = - \frac{\delta_{ij} r_a^{\alpha}}{\alpha}
		+ \frac{\pa^a_i \pa^a_j r_a^{\alpha+2}}{\alpha (\alpha + 2)}\,, \\
	r_a^{\alpha} n_a^i n_a^j n_a^k =
		\frac{( \delta_{ij} \pa^a_k + \delta_{ik} \pa^a_j
		+ \delta_{jk} \pa^a_i ) r_a^{\alpha+1}}{(\alpha - 1)(\alpha + 1)}
		- \frac{\pa^a_i \pa^a_j \pa^a_k r_a^{\alpha+3}}
			{(\alpha - 1)(\alpha + 1)(\alpha + 3)}\,.
\end{gather}
These equations are sometimes not defined without regularization, because the
right-hand side might diverge in special cases. The derivatives $\pa^1_i $ and
$\pa^2_j$ are now pulled out in front of the integral, and we can use the
famous formula from \cite{R49} to carry out the integrations:
\begin{equation}
\left[\int d^3x\,r^{\alpha}_{1}r^{\beta}_{2}\right]_{\text{reg}} \equiv
	\pi^{3/2}\frac{
		\Gamma\left(\frac{\alpha+3}{2}\right)\Gamma\left(\frac{\beta+3}{2}\right)
		\Gamma\left(-\frac{\alpha+\beta+3}{2}\right)}
	{\Gamma\left(-\frac{\alpha}{2}\right)\Gamma\left(-\frac{\beta}{2}\right)
		\Gamma\left(\frac{\alpha+\beta+6}{2}\right)}
	r_{12}^{\alpha+\beta+3}\,.
\end{equation}

After the partial derivatives $\pa^1_i $ and $\pa^2_j$ with respect to
the particle positions, that were pulled out of the integral before, are performed,
the limit $\epsilon \rightarrow 0$ can be considered. In all cases emerging in our
calculations this limit was independent of the direction in the $(\mu, \nu)$-plane.
Astonishingly, the most complicated integral appearing in this work has the simplest
solution:
\be
	\int d^3 x \, h^{\text{TT} \, \text{spin}}_{(4) i j}
		\phi_{(2) , i} \phi_{(2) , j} = 0\,.
\ee

\end{document}